\documentclass[%
 reprint,
 twocolumn,
 amsmath,amssymb,
 aps,
 prl,
 nofootinbib,
 tightenlines,
 nobibnotes,
 longbibliography
]{revtex4-2}

\usepackage{graphicx}
\usepackage{dcolumn}
\usepackage{bm}
\usepackage{hyperref}

\usepackage{color}
\usepackage{mathrsfs}
\usepackage{amsmath}
\usepackage{stfloats}
\usepackage{float}
\usepackage{gensymb}
\usepackage{braket}

\definecolor{Green}{RGB}{0,100,0}
\definecolor{Blue}{RGB}{51,153,255}
\definecolor{Red}{RGB}{151,010,010}

\usepackage{capt-of}

\begin{document}

\title{
Phase-resolved measurement of entangled states via common-path interferometry
}

\author{Andrew A. Voitiv$^1$}
\author{Mark T. Lusk$^2$}%
\email[]{mlusk@mines.edu}%
\author{Mark E. Siemens$^1$}
\email[]{msiemens@du.edu}%
\affiliation{$^1$Department of Physics and Astronomy, University of Denver, Denver, CO 80208, USA
}%
\affiliation{
$^2$Department of Physics, Colorado School of Mines, Golden, CO 80401, USA
}%

\date{\today}

\begin{abstract}
We propose and experimentally demonstrate a method to directly measure the phase of biphoton states using an entangled mode as a collinear reference. The technique is demonstrated with entangled photonic spatial modes in the Laguerre-Gaussian basis, and it is applicable to any pure quantum system containing an exploitable reference state in its entanglement spectrum.  As one particularly useful application, we use the new methodology to directly measure the geometric phase accumulation of entangled photons.
\end{abstract}

\maketitle

\section{Introduction}

\begin{figure} [h!]
\centering
\includegraphics[width=.95\linewidth]{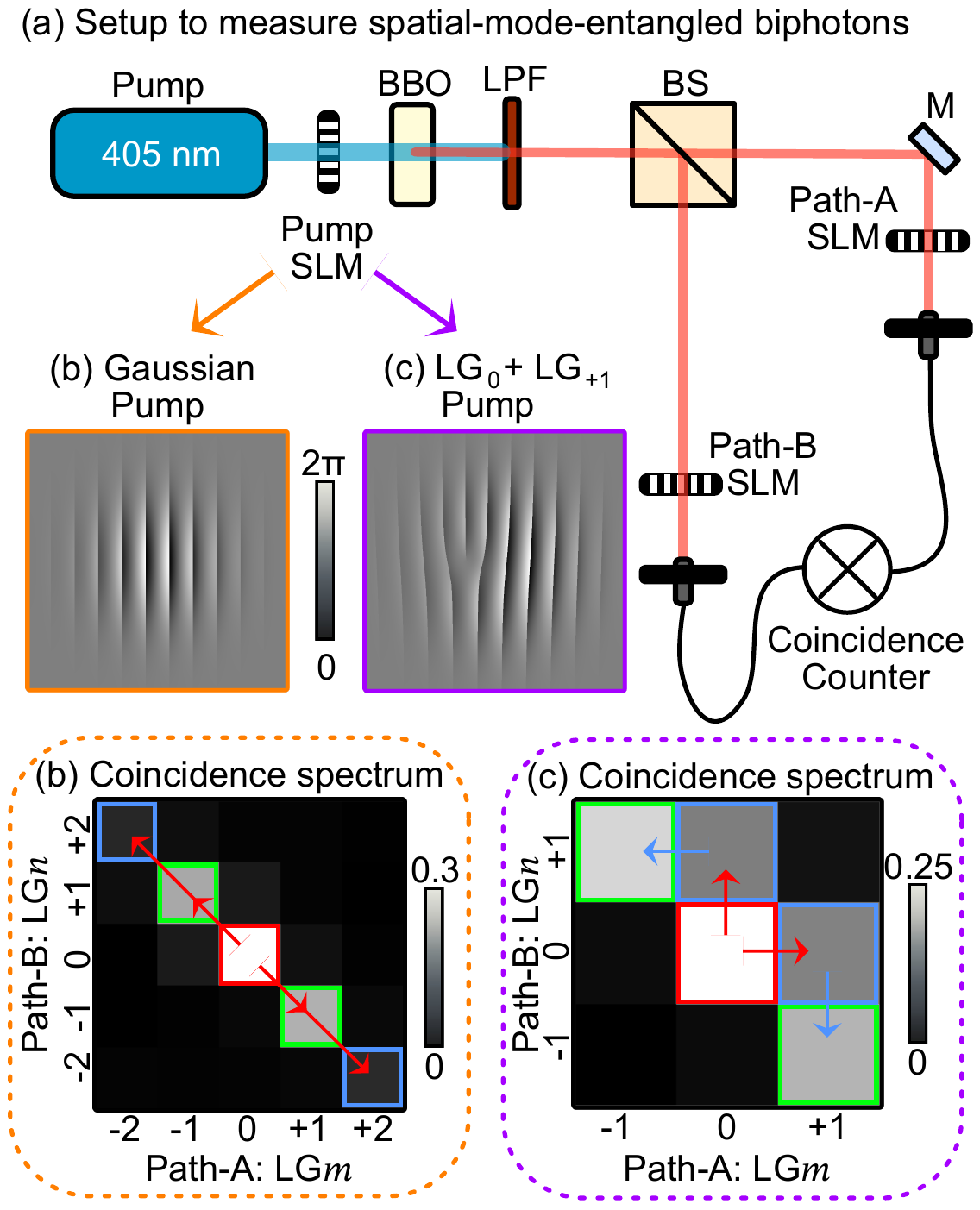}
\caption{(a) Simplified schematic of entangled biphoton coincidence measurements. BBO: Type-I beta barium-borate nonlinear crystal; LPF: longpass filter; BS: $50/50$ beamsplitter; M: mirror; SLM: spatial light modulator. (b) Experimental coincidence spectrum from a Gaussian laser pump in the $\mathrm{LG}^{p=0}_{m}$ basis. The first experimental application uses the zero-order correlated state (red square) as an ancillary collinear reference mode to make phase measurements on higher-order states, such as those highlighted in green and blue squares. (c) Experimental coincidence spectrum from a pump beam that is a superposition of $\mathrm{LG}_{0}$ and $\mathrm{LG}_{+1}$ modes. Here, the phase of each entangled mode can be detected by inductive relative phase measurements (red square to blue and then blue to green).}
\label{fig:cartoonschematic}
\end{figure}

Entangled photons are a valuable resource for quantum logic, imaging, and information theory \cite{Bouwmeester1998, Kim2000, Couteau2018, Proietti2019, Magana-Loaiza2019}.  While measuring the amplitude of entangled states is relatively straightforward with coincidence-based correlation filters \cite{Kwiat1995, Mair2001, Romero2011, Mclaren2012,McLaren2015, Restuccia2016, Baghdasaryan2020, Lib2020, Valencia2021}, such as shown in Fig. \ref{fig:cartoonschematic}, the corresponding measurement of entangled state phases has received relatively little attention. In the classical treatment of light, phase can be easily measured using interference with an ancillary reference beam \cite{Picart2015}. Biphoton measurements rely on coincidence detection, though, and the requisite reference is no longer straightforward. Past approaches to measuring entangled quantum states have relied on a series of projective measurements, with orthonormal basis states and superpositions thereof, to indirectly compute the elements of a state operator by fitting those measurements to a predicted model (and iterating the model to minimize errors in the reconstructed state) \cite{Agnew2011,Toninelli2019, Li2023, Sahoo2020}. More recently, a new experimental technique has emerged in which an additional degree of  entanglement is introduced to facilitate a reference for measuring phase \cite{Pan2019}. This approach is experimentally complex and restricted to photons in the polarization basis. This motivates a search for simpler and more generically applicable ways for measuring the phase of entangled states, especially in high-dimensional systems.

In this paper, we demonstrate the use of an  entangled mode as a collinear reference for direct measurement of pure multi-state quantum systems. By \textit{direct measurement}, we mean that this method uses only experimental data and does not require the indirect computation employed in traditional quantum state tomography. By treating one of the constituent components in the entangled state as a reference mode, collinear interference can be measured without any alteration to the standard setup of biphoton coincidence measurements. Phase is retrieved from this interference by phase shifting the measurement apparatus, rather than the reference mode itself, which reduces the need for any additional optical components. Furthermore, this method bypasses the need to identify requisite orthonormal basis states, required in quantum state tomography; all information needed on how to perform phase measurements comes directly from simple coincidence measurements in a single basis. The Laguerre-Gaussian (LG) basis of entangled biphotons is used to both introduce and apply the measurement strategy. In addition to demonstrating the full experimental characterization of entangled state vectors, we apply the approach to directly measure the geometric phase accumulated by biphotons that propagate through misaligned Dove prisms. We anticipate that this method can be readily integrated into any coincidence-measurement experiment for which at least one component of a pure state can be used as a reference.

\section{Concept of Measuring Phase of Entangled States} \label{Theory}

Consider a two-particle entangled state of quantum systems $A$ and $B$. Define a basis of $\{\ket{m}_A\}$ for the first system and $\{\ket{n}_B\}$ for the second. The entangled state is then
\begin{equation} \label{full}
    \ket{\Psi} = \sum c_{m,n} \ket{m}_A \ket{n}_B \equiv \sum c_{m,n} \ket{m,n},
\end{equation}
with amplitudes $a_{m,n}$ and phases $\varphi_{m,n}$ comprising the complex coefficients, $c_{m,n} = a_{m,n} e^{i \varphi_{m,n}}$.

It is common to measure the amplitude of an entangled component (or \textit{mode}) of the composite system by performing projective measurements tuned to measure the mode in question on both signal and idler arms of the experiment and then recording coincidences counts between them. This is represented with the following measurement device:
\begin{equation} \label{devices}
        \ket{\Phi^{\mathrm{Device}} }= \ket{\Phi}_A \ket{\Phi}_B = \ket{p,q},
\end{equation}
where $\ket{\Phi}_A = \ket{p}_A$ and $\ket{\Phi}_B = \ket{q}_B$.
A projective measurement of mode amplitude is then 
\begin{equation}
a_{p,q}^2 = |\braket{\Phi | \Psi}|^2.
\end{equation}
The \textit{coincidence} entanglement spectrum is obtained by carrying out such  measurements for all modes. Two such spectra are shown in Fig. \ref{fig:cartoonschematic}.

We turn now to measurement of the phase of each of the modes in the entanglement spectrum for which, to the best of our knowledge, there is no established direct method. Our approach is inspired by classical wave dynamics, in which interference with a reference wave is used to measure phase. The difference is that we do not introduce an external reference; rather, we select and interfere specific modes from the broader entanglement spectrum. Thus, no external systems or multi-arm interferometry are required.

The phase of the coefficient on a mode, $\ket{m=p,n=q}$, in Eqn. \ref{full} can be determined as follows. Construct a measurement device that probes the interference between $\ket{p,q}$ and a finite amplitude reference, $\ket{p_{\mathrm{ref}},q_{\mathrm{ref}}}$, in the entangled state:
\begin{equation} \label{device_reference}
\ket{\Phi^{\mathrm{Reference}}} = \ket{p,q} + e^{i \varphi_{\mathrm{ref}}} d_{\mathrm{ref}} \ket{p_{\mathrm{ref}}, q_{\mathrm{ref}}}
\end{equation}
where $d_{\mathrm{ref}}$ controls the reference mode strength to optimize interference contrast, and $\varphi_{\mathrm{ref}}$ shifts the reference phase. 

The prescription of which reference mode to use is chosen such that the projection measurement is in the form of two-wave interference:
\begin{equation} \label{interference}
|\braket{\Phi | \Psi}|^2 = |  c_{p,q} + c_{p_{\mathrm{ref}},q_{\mathrm{ref}}} d_{\mathrm{ref}} e^{-i \varphi_{\mathrm{ref}}}|^2.
\end{equation}
This interferogram is solvable for the phase of the coefficient $c_{p,q}$, relative to the reference phase of $c_{p_{\mathrm{ref}},q_{\mathrm{ref}}}$. Shortly, we discuss an example in choosing these reference modes.

A common problem in classical physics is solving for the sign of the phase from two-wave interference, due to a quadrature ambiguity for the inverse cosine function, and many solutions have been identified \cite{Picart2015}. A simple approach is based on phase-shifting digital holography (PSDH) \cite{Yamaguchi1997},  a technique that uses separate interference measurements, each with different phase shifts applied to the reference wave. The most common implementation is to make four interference measurements, with the reference phase, $\varphi_{\mathrm{ref}}$, stepped by $\pi/2$ between each, and combine them into
an arctangent function to compute the phase with the correct sign:
\begin{equation} \label{PSDH}
    \varphi = - \textrm{arctan}{\left(I_{\pi/2} - I_{3\pi/2},I_{0} - I_{\pi} \right)}.
\end{equation}
Here arctan$(y,x)$ has an unambiguous sign but is otherwise equivalent to $\arctan(y/x)$. 

PSDH can be applied to a bipartite quantum system in reverse. Instead of making and actively phase shifting the reference state \cite{Andersen2019}, we measure for it by phase shifting $\varphi_{\mathrm{ref}}$ in Eqn. \ref{interference}. By making four phase-stepped coincidence measurements, as prescribed in steps of $\pi/2$, and then combining them into Eqn. \ref{PSDH}, the phase of interest, $\varphi_{p,q}$, can be retrieved.

The entire phase spectrum can be determined by designating one mode as the absolute reference, measuring the phase of its nearest neighbors, and then using those modes as relative references to measure the phase of their neighbors. For example, the phase of $\ket{p=1,q=0}$ can be measured by constructing the coincidence devices, that build Eqn. \ref{device_reference}, with $q_{\mathrm{ref}} = q = 0$ and $p_{\mathrm{ref}} = 0$:
\begin{equation} \label{(1,0)}
    \ket{\Phi}_A = \ket{1}_A + e^{i \varphi_{\mathrm{ref}}} d_{\mathrm{ref}} \ket{0}_A, \, \, \ket{\Phi}_B = \ket{0}_B
\end{equation}
which constitute the coincidence device, $\ket{\Phi^{(1,0)}} = \ket{\Phi}_A \ket{\Phi}_B$. With this, one can measure four coincidence readings (Eqn. \ref{interference}) of different phase shifts, $\varphi_{\mathrm{ref}}$ in steps of $\pi/2$, to be combined in Eqn. \ref{PSDH} to compute the phase $\varphi_{1,0}$. The phase $\varphi_{0,1}$ is similarly measured:
\begin{equation} \label{(0,1)}
    \ket{\Phi}_A = \ket{0}_A, \, \, \ket{\Phi}_B = \ket{1}_B + e^{i \varphi_{\mathrm{ref}}} d_{\mathrm{ref}} \ket{0}_B
\end{equation}
which constitute $\ket{\Phi^{(0,1)}} = \ket{\Phi}_A \ket{\Phi}_B$. In both cases, the reference phase is applied to the path-separated device for which an interference of single-system modes is being measured.

Continuing the same example, the phase of modes that are nearest neighbors to $\ket{0,1}$ and $\ket{1,0}$ can be subsequently evaluated by using these phase-measured modes as new references. For instance, choosing either one can obtain the phase of mode $\ket{1,1}$, relative to the chosen reference. In general, an \textit{inductive} procedure can be followed whereby the phase of state $\ket{m,n}$ is sufficient to evaluate the phase of states $\ket{m\pm 1, n\pm 1}$. A phase spectrum of any bandwidth can be measured in this way, based on the choice of a preliminary mode to serve as the reference gauge.

If, however, the chosen reference mode does not have any nearest neighbors of finite amplitude, then the procedure above can be followed but with next-nearest (diagonal) neighbors. Alternatively, an \textit{ancillary} reference mode may be identifiable in a spectrum. If a mode $\ket{p,q}$ has finite amplitude and all other states containing either $\ket{p}$ or $\ket{q}$ have zero amplitude, then that isolated state can be used as the sole reference to measure the phase of the full selected spectrum. Such ancillary cases are rare, but one example is shown in  the single-diagonal spectrum of Fig. \ref{fig:cartoonschematic} (a); subpanel (b) shows a more general example for which the inductive approach is required.

In the next three sections, we apply this methodology to experimentally measure the phase of entangled states associated with correlated biphotons from spontaneous parametric downconversion (SPDC). We begin with the particularly simple spectrum of Fig. \ref{fig:cartoonschematic} (a), amenable to the ancillary approach. We then use that as a setting to make direct measurements of geometric phases accumulated by the biphoton system under propagation through Dove prisms in one path. As a second application, we work with the more complicated spectrum of Fig. \ref{fig:cartoonschematic} (b), to experimentally implement the more general inductive method.

\section{Ancillary reference for measuring phase of entangled modes} \label{Application1}

The new methodology was first applied to directly measure the phase of correlated photon pairs exiting a laser-pumped nonlinear crystal, $\beta$ barium-borate cut for Type-I SPDC \cite{Mair2001}. The biphoton state can be expressed as a superposition of Laguerre-Gaussian (LG) modes for each photon, $A$ and $B$, in a form identical to Eqn. \ref{full}, with $\ket{m} = \mathrm{LG}^{p=0}_m$ for topological charge $m$ and with attention restricted to radial mode order $p=0$. When a Gaussian pump photon of zero orbital angular momentum (OAM) interacts with the nonlinear crystal, the down-converted biphotons are correlated with $n=-m$ in Eqn. \ref{full} because angular momentum is conserved \cite{Saleh2000, Mair2001}. This setting is depicted in the simplified schematic of Fig. \ref{fig:cartoonschematic}. See \S S.I of the Supplementary Document for a full schematic and complete details of the experimental apparatus, including the definition for LG mode spatial profiles.

It is common to measure the amplitude of a particular mode of the entangled spectrum by programming LG-mode filters on the two paths with spatial light modulators (SLMs); by using the complex conjugate of an OAM mode on each SLM, the first-diffracted orders are subsequently phase-\textit{flattened} and thus are coupled into single-mode optical fibers \cite{Bouchard2018}. For example, to measure the amplitude of there being the mode $\ket{m}$ in the $A$-path and $\ket{n}$ in the $B$-path in Eqn. \ref{full}, each SLM is programmed according to Eqn. \ref{devices}:
\begin{equation}
\begin{gathered}
        \ket{\Phi^{\mathrm{SLM}}}_A = \ket{m}_A, \, \ket{\Phi^{\mathrm{SLM}}}_B = \ket{n}_B, \\
        \ket{\Phi^{\mathrm{SLM}} }= \ket{\Phi}_A \ket{\Phi}_B = \ket{m,n}.
\end{gathered}
\end{equation}
Performing a series of these measurements for different pairs of $\ket{m}_A$ and $\ket{n}_B$ results in a coincidence spectrum as seen in Fig. \ref{fig:cartoonschematic} (a).

\begin{figure} [h!]
\centering
\includegraphics[width=.99\linewidth]{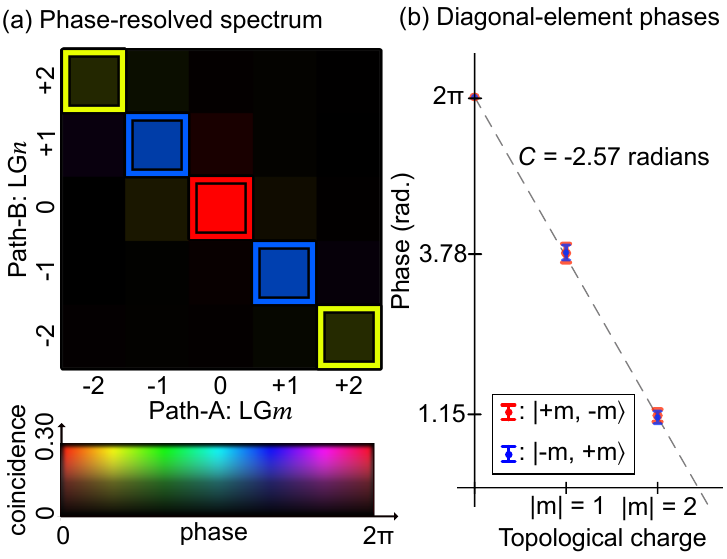}
\caption{(a) Experimental measurements of coincidences ($a_{m,n}^2$) and phases ($\varphi_{m,n}$) of entangled LG modes up to order $m = 2$. The intensities (coincidences) are normalized such that all entries sum to one. Numerical values for the main diagonal coincidences and phases are tabulated in \S S.3 the Supplementary Document. All coincidence measurements (including those used to calculate the phase) are the averages of five coincidence measurements taken every 10 seconds. Main-diagonal elements are boxed in their corresponding color of the phase legend; the phase of the center mode, $\ket{0,0}$, is set to zero by definition as the reference. (b) Phases of the main-diagonal elements. Error bars for the phases were calculated by propagating the standard deviations of the four different measurements required \cite{Taylor1997}. The dashed line is a fit of Eq. \ref{relative} to the data points, yielding $C = -2.57$ radians.}
\label{fig:spectrumplot}
\end{figure}

We move from these amplitude measurements to phase measurements of the modes with finite amplitude ($n=-m$) using the $\ket{0,0}$ mode as ancillary reference. The SLMs are program as follows to define a coincidence measurement $\ket{\Phi^{\mathrm{SLM}}} = \ket{\Phi}_A \ket{\Phi}_B$:
\begin{equation}
\begin{gathered}
        \ket{\Phi^{\mathrm{SLM}}}_A = d_m \ket{m}_A + e^{i \varphi_{\mathrm{ref}}} \, d_0 \ket{0}_A  \\
        \ket{\Phi^{\mathrm{SLM}}}_B = d_m \ket{-m}_B + d_0 \ket{0}_B.
\end{gathered}
\end{equation}
As described above, $\varphi_{\mathrm{ref}}$ is to be incremented in four steps of $\pi/2$, prescribed by Eqn. \ref{PSDH}. The resultant interference is:
\begin{equation}
    |\braket{\Phi|\Psi}|^2 = |d_m^{2} a_m e^{i \varphi_{\pm}^{\scriptscriptstyle (m)}} + d_0^{2} a_0 e^{i(\varphi_{0} - \varphi_{\mathrm{ref}})}|^2.
\end{equation}
Without loss of generality, set $\varphi_{0} = 0$ because all phase measurements are defined relative to this ancillary zero-order mode. By taking four phase-stepped coincidence readings of the above interference, the phase $\varphi_{\pm}^{\scriptscriptstyle (m)}$ is calculated via Eqn. \ref{PSDH}. Our phase-resolved experimental measurements are shown in Fig. \ref{fig:spectrumplot} (a), for a subset of Eqn. \ref{full} up to $|m|=2$ with $n=-m$; the brightness of that plot corresponds with typical coincidence measurements, just as seen in Fig. \ref{fig:cartoonschematic} (a). The color gradient of that plot represents the new phase measurements of each component of the spectrum. \S S.1 of the Supplementary Document contains the values used for $d_0$, $d_1$, and $d_2$ for the measurement gratings.

The experimental spectrum of Fig. \ref{fig:spectrumplot}  
shows a coincidence character consistent with angular momentum conservation for a simple Gaussian pump \cite{Saleh2000}. 
It is the measured phase variation along the diagonal, though, that provides a crucial test of our methodology. Propagation of a single-photon LG mode, $m$, through the optical path generates a Gouy phase proportional to the magnitude of its mode number,
\begin{equation} \label{gouy}
    \varphi_{\mathrm{Gouy},m} = -(|m| + 1) \arctan{\frac{z}{z_R}},
\end{equation}
where Rayleigh length $z_R = \pi w_0^2 / \lambda$. For fixed values of wavelength ($\lambda$), beam waist ($w_0$), and propagation distance ($z$), this phase will vary only with $m$. The other factors comprise a constant.
This implies that biphoton modes, $\ket{\pm m}_A \ket{\mp m}_B$, will accumulate a two-photon Gouy phase, relative to the phase of the  ancillary ($m=0$) mode, that is the sum of the Gouy phases of the associated single-photon transits and scales linearly with $|m|$:
\begin{equation} \label{relative}
\varphi_{|m|} = \varphi_{\mathrm{Gouy},(\pm m,\mp m)} - \varphi_{\mathrm{Gouy},(0,0)} = C \, |m|.
\end{equation}
The constant, $C$, can therefore be determined from the fitted slope of the change in phase along the diagonal and was found to be $-2.57$ radians. Fig. \ref{fig:spectrumplot} (b) shows the measured phases for $|m| = 1$ and $2$, plotted together, while Supplementary Document \S S.5 provides data up to $|m|=3$ for a different setting of $w_0$ on the measurement SLMs. While one would expect the Gouy phase to be zero for an ideal optical system, in our experiment we hypothesize the nonzero phase shifts arise from the optical planes not being aligned precisely for each of the different lenses used for both photon paths \cite{Agnew2011}---see \S S.1 for a full schematic and description of the experimental components.

The data of Fig. \ref{fig:spectrumplot} (a) can be used to directly construct the full quantum state using only experimental measurements:
\begin{equation} \label{reconstructed}
    \ket{\Psi^{\mathrm{experiment}}} = \sum_{m,n} a_{m,n} e^{i \varphi_{m,n}} \ket{m,n}.
\end{equation}
The density matrix representation of this state is visualized in \S S.4 of the Supplementary Document.

\section{Ancillary reference for measuring geometric phase shifts of entangled states}

A second application demonstrates  quantitatively accurate sensing of controlled changes of the entangled state phase. We inserted two Dove prisms in the $A$-path to add geometric phase that can be varied by changing their relative orientation. With a Gaussian pump beam, the final $m=1$ state of the system can be anticipated using a result from classical optics \cite{Padgett1999, Galvez2003, Lusk2022}:
\begin{equation} \label{first_order_geometric}
    \ket{\Psi^{\mathrm{final}}} = \frac{1}{\sqrt{2}}\bigl(e^{i 2\eta} \ket{+1,-1} + e^{-i 2\eta} \ket{-1,+1}\bigr).
\end{equation}
Here $\eta$ is the angle of misorientation of the prisms, and the argument, $2 \eta$, is the geometric phase shift \cite{Galvez2003}.

\begin{figure} [h!]
\centering
\includegraphics[width=.99\linewidth]{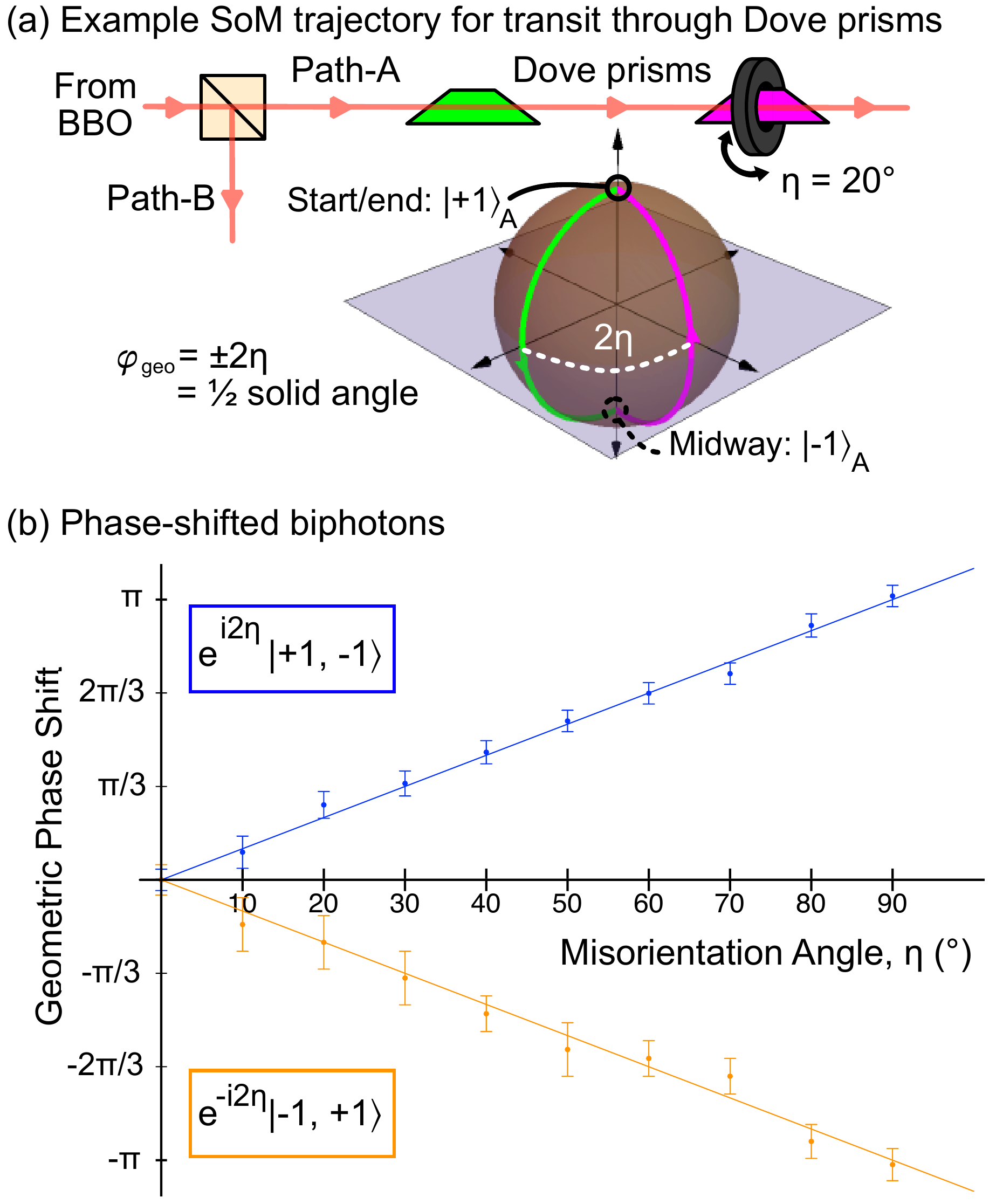}
\caption{(a) An example trajectory of a photon on the first-order Sphere of Modes (SoM) \cite{Lusk2022}. (b) Measured geometric phase shifts of entangled biphotons after transmission through a pair of Dove prisms with misorientation angle $\eta$. All data points are reported relative to the $\eta = 0\degree$ point for each set. The flip in sign of the geometric phase is explained by the different handedness of the trajectories taken by the photons on the SoM \cite{Voitiv2022}. Density matrix representations of a few of these phase-shifted states are plotted in \S S.6 of the Supplementary Document.}
\label{fig:geometricphase}
\end{figure}

Fig. \ref{fig:geometricphase} (a) shows a representative trajectory on the Sphere of Modes (SoM), the OAM analog of the polarization Poincar\'e sphere \cite{Padgett1999}, with the geometric phase identifiable as one-half the solid angle enclosed. Panel (b) compares the experimentally measured modal phase shifts with the anticipated geometric phase  for several different misorientation angles. Furthermore, we confirmed that our method works by phase shifting the path-$B$ SLM rather than the path-$A$ SLM, even though the Dove prisms are only in path-$A$; these results are presented in \S S.7 of the Supplementary Document. This demonstrated versatility in choosing how to phase step the reference mode may be useful in quantum sensing and imaging, in settings where only the idler-path measurement device were amenable to user control for measuring the phase accumulated in the signal path.

\section{Inductive Phase Measurements of Entangled Modes}

\begin{figure} [h!]
\centering
\includegraphics[width=.99\linewidth]{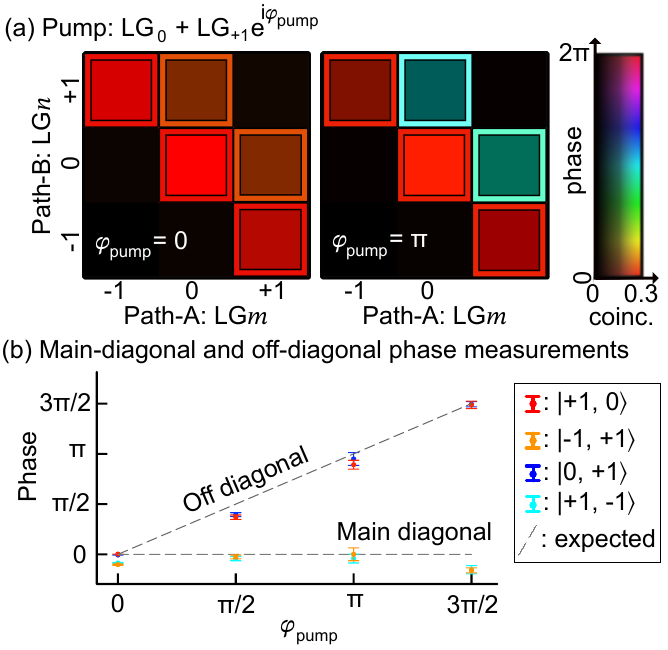}
\caption{(a) Phase-resolved coincidence spectra for two pump beams of different phase shifts between $\mathrm{LG}_0$ and $\mathrm{LG}_{+1}$ modes. For an applied shift of zero, the entangled biphotons are measured to have a near-zero shift between main-diagonal modes (contributed from $\mathrm{LG}_0$) and off-diagonal modes (contributed from $\mathrm{LG}_{+1}$). In the second case, a $\pi$ phase shift is applied. There, the intensity of the coincidence spectrum is unaltered (within small experimental fluctuations) but the $\pi$ phase-shift between the main- and off-diagonal modes is clearly present. (b) Measured phases of the four finite amplitudes in the spectra, including error bars using the same methods described above, resulting from four different phase-shifted pump beams. Dashed lines are the predicted shift one would expect to measure in the biphotons based on the shift applied to the pump.
}
\label{fig:structuredpump}
\end{figure}

Lastly, we provide a demonstration of measuring the phase of entangled photons when there is not an available ancillary mode, as seen in Fig. \ref{fig:cartoonschematic} (b). To do so, the Gaussian pump beam is replaced by to one structured with an SLM before transiting the BBO. (See \S S.8 of the Supplementary Document for the complete schematic.) We generated a superposition of a fundamental Gaussian and one containing an optical vortex: $\mathrm{LG}_0 + \mathrm{LG}_{+1} \, e^{i \, \varphi_{\mathrm{pump}}}$, where $\varphi_{\mathrm{pump}}$ is a controllable phase shift on the pump vortex. The biphoton state resulting from such a pump has long-standing predictions \cite{Franke-Arnold2002, Bornman2021}, and we present consistent phase-resolved coincidence spectra of our experimental measurements in Fig. \ref{fig:structuredpump}. It is evident that there is a main-diagonal from the $\mathrm{LG}_0$ contribution of the pump and an off-diagonal from the $\mathrm{LG}_{+1}$ contribution.

To measure those phase-resolved spectra using the \textit{inductive} approach, we first used $\ket{0,0}$ as reference mode to measure both $\ket{0,+1}$ and $\ket{+1,0}$ separately, as described above in defining Eqns. \ref{(0,1)} and \ref{(1,0)}, respectively. From there, the inductive method is applied using these two measured states as the new reference modes. To measure $\ket{+1,-1}$, we used $\ket{+1,0}$ as the reference, phase-stepping the path-$B$ SLM. To measure $\ket{-1,+1}_B$, we used $\ket{0,+1}$ as the reference, phase-stepping the path-$A$ SLM.

Fig. \ref{fig:structuredpump} clearly shows that phase shifts are measured between diagonal and off-diagonal modes with the inductive method. These phase shifts were generated by simply changing the phase, $\varphi_{\mathrm{pump}}$, in the hologram on the SLM that structured the pump beam. The versatility of this inductive approach to quantum state measurement can be readily applied to other arbitrarily-produced and structured entanglement spectra.

\section{Conclusion}

We have proposed and experimentally demonstrated a method to directly measure the phase of biphoton states using an ancillary entangled mode. The methodology was validated using entangled Laguerre-Gaussian spatial modes, but it is applicable to any pure quantum system that supports an entangled ancillary mode. This allows for a purely experimental approach to measure a quantum state without reliance on the indirect computational methods employed in quantum state tomography. In contrast to these existing methods, ours is direct because we employ analytical relationships between specific intermodal interferences to determine phase, thus bypassing the need to fit many projective measurements to a model. Our direct phase methods could find application for characterizing systems with an unknown number of mutually unbiased basis states, which is a challenge for computational-based quantum state tomography. 

Our theoretical proposal, and experimental demonstration, considered only pure entangled states (as we selected one mode to measure relative phases of other modes in a state). But given a pure state, our approach is not limited by the dimensionality of the system: if the amplitude of a mode can be measured, then so too can its phase with the additional interferometric measurements. It is also possible that this tomographic technique could be applied in conjunction with phase measurements of N00N states that are used for high-precision parameter estimation (quantum metrology) \cite{Hiekkamaki2021, Hong2021, Nielsen2023}. Lastly, our methodology was used to directly measure geometric phase accumulation, demonstrating the ability to quantitatively and accurately sense changes of inter-mode entangled state phases.

\textbf{Supplementary Material.} A Supplementary Document is provided which contains full details on the complete experimental schematic and methods, along with additional applications of the new technique.

\textbf{Acknowledgements.} The authors thank Z. Kan for early assistance with our experimental methods and P. C. Ford for initial SLM and hologram support.

\textbf{Funding.} W. M. Keck Foundation.

\textbf{Disclosures.} The authors declare no conflicts of interest.

\textbf{Data availability.} Data underlying all results presented are available from the authors upon reasonable request.

\bibliography{Refs.bib}

\begin{thebibliography}{34}%
\makeatletter
\providecommand \@ifxundefined [1]{%
 \@ifx{#1\undefined}
}%
\providecommand \@ifnum [1]{%
 \ifnum #1\expandafter \@firstoftwo
 \else \expandafter \@secondoftwo
 \fi
}%
\providecommand \@ifx [1]{%
 \ifx #1\expandafter \@firstoftwo
 \else \expandafter \@secondoftwo
 \fi
}%
\providecommand \natexlab [1]{#1}%
\providecommand \enquote  [1]{``#1''}%
\providecommand \bibnamefont  [1]{#1}%
\providecommand \bibfnamefont [1]{#1}%
\providecommand \citenamefont [1]{#1}%
\providecommand \href@noop [0]{\@secondoftwo}%
\providecommand \href [0]{\begingroup \@sanitize@url \@href}%
\providecommand \@href[1]{\@@startlink{#1}\@@href}%
\providecommand \@@href[1]{\endgroup#1\@@endlink}%
\providecommand \@sanitize@url [0]{\catcode `\\12\catcode `\$12\catcode `\&12\catcode `\#12\catcode `\^12\catcode `\_12\catcode `\%12\relax}%
\providecommand \@@startlink[1]{}%
\providecommand \@@endlink[0]{}%
\providecommand \url  [0]{\begingroup\@sanitize@url \@url }%
\providecommand \@url [1]{\endgroup\@href {#1}{\urlprefix }}%
\providecommand \urlprefix  [0]{URL }%
\providecommand \Eprint [0]{\href }%
\providecommand \doibase [0]{https://doi.org/}%
\providecommand \selectlanguage [0]{\@gobble}%
\providecommand \bibinfo  [0]{\@secondoftwo}%
\providecommand \bibfield  [0]{\@secondoftwo}%
\providecommand \translation [1]{[#1]}%
\providecommand \BibitemOpen [0]{}%
\providecommand \bibitemStop [0]{}%
\providecommand \bibitemNoStop [0]{.\EOS\space}%
\providecommand \EOS [0]{\spacefactor3000\relax}%
\providecommand \BibitemShut  [1]{\csname bibitem#1\endcsname}%
\let\auto@bib@innerbib\@empty
\bibitem [{\citenamefont {Bouwmeester}\ \emph {et~al.}(1997)\citenamefont {Bouwmeester}, \citenamefont {Pan}, \citenamefont {Mattle}, \citenamefont {Eibl}, \citenamefont {Weinfurter},\ and\ \citenamefont {Zeilinger}}]{Bouwmeester1998}%
  \BibitemOpen
  \bibfield  {author} {\bibinfo {author} {\bibfnamefont {D.}~\bibnamefont {Bouwmeester}}, \bibinfo {author} {\bibfnamefont {J.-W.}\ \bibnamefont {Pan}}, \bibinfo {author} {\bibfnamefont {K.}~\bibnamefont {Mattle}}, \bibinfo {author} {\bibfnamefont {M.}~\bibnamefont {Eibl}}, \bibinfo {author} {\bibfnamefont {H.}~\bibnamefont {Weinfurter}},\ and\ \bibinfo {author} {\bibfnamefont {A.}~\bibnamefont {Zeilinger}},\ }\bibfield  {title} {\bibinfo {title} {{Experimental quantum teleportation}},\ }\href@noop {} {\bibfield  {journal} {\bibinfo  {journal} {Nature}\ }\textbf {\bibinfo {volume} {390}} (\bibinfo {year} {1997})}\BibitemShut {NoStop}%
\bibitem [{\citenamefont {Kim}\ \emph {et~al.}(2000)\citenamefont {Kim}, \citenamefont {Yu}, \citenamefont {Kulik},\ and\ \citenamefont {Shih}}]{Kim2000}%
  \BibitemOpen
  \bibfield  {author} {\bibinfo {author} {\bibfnamefont {Y.-H.}\ \bibnamefont {Kim}}, \bibinfo {author} {\bibfnamefont {R.}~\bibnamefont {Yu}}, \bibinfo {author} {\bibfnamefont {S.~P.}\ \bibnamefont {Kulik}},\ and\ \bibinfo {author} {\bibfnamefont {Y.}~\bibnamefont {Shih}},\ }\bibfield  {title} {\bibinfo {title} {{Delayed "Choice" Quantum Eraser}},\ }\href@noop {} {\bibfield  {journal} {\bibinfo  {journal} {Physical Review Letters}\ }\textbf {\bibinfo {volume} {84}},\ \bibinfo {pages} {1} (\bibinfo {year} {2000})}\BibitemShut {NoStop}%
\bibitem [{\citenamefont {Couteau}(2018)}]{Couteau2018}%
  \BibitemOpen
  \bibfield  {author} {\bibinfo {author} {\bibfnamefont {C.}~\bibnamefont {Couteau}},\ }\bibfield  {title} {\bibinfo {title} {{Spontaneous parametric down-conversion}},\ }\href@noop {} {\bibfield  {journal} {\bibinfo  {journal} {Contemporary Physics}\ }\textbf {\bibinfo {volume} {59}},\ \bibinfo {pages} {3} (\bibinfo {year} {2018})}\BibitemShut {NoStop}%
\bibitem [{\citenamefont {Proietti}\ \emph {et~al.}(2019)\citenamefont {Proietti}, \citenamefont {Pickston}, \citenamefont {Graffitti}, \citenamefont {Barrow}, \citenamefont {Kundys}, \citenamefont {Branciard}, \citenamefont {Ringbauer},\ and\ \citenamefont {Fedrizzi}}]{Proietti2019}%
  \BibitemOpen
  \bibfield  {author} {\bibinfo {author} {\bibfnamefont {M.}~\bibnamefont {Proietti}}, \bibinfo {author} {\bibfnamefont {A.}~\bibnamefont {Pickston}}, \bibinfo {author} {\bibfnamefont {F.}~\bibnamefont {Graffitti}}, \bibinfo {author} {\bibfnamefont {P.}~\bibnamefont {Barrow}}, \bibinfo {author} {\bibfnamefont {D.}~\bibnamefont {Kundys}}, \bibinfo {author} {\bibfnamefont {C.}~\bibnamefont {Branciard}}, \bibinfo {author} {\bibfnamefont {M.}~\bibnamefont {Ringbauer}},\ and\ \bibinfo {author} {\bibfnamefont {A.}~\bibnamefont {Fedrizzi}},\ }\bibfield  {title} {\bibinfo {title} {{Experimental test of local observer independence}},\ }\href@noop {} {\bibfield  {journal} {\bibinfo  {journal} {Science Advances}\ }\textbf {\bibinfo {volume} {5}},\ \bibinfo {pages} {9} (\bibinfo {year} {2019})}\BibitemShut {NoStop}%
\bibitem [{\citenamefont {Maga{\~{n}}a-Loaiza}\ and\ \citenamefont {Boyd}(2019)}]{Magana-Loaiza2019}%
  \BibitemOpen
  \bibfield  {author} {\bibinfo {author} {\bibfnamefont {O.~S.}\ \bibnamefont {Maga{\~{n}}a-Loaiza}}\ and\ \bibinfo {author} {\bibfnamefont {R.~W.}\ \bibnamefont {Boyd}},\ }\bibfield  {title} {\bibinfo {title} {{Quantum imaging and information}},\ }\href@noop {} {\bibfield  {journal} {\bibinfo  {journal} {Reports on Progress in Physics}\ }\textbf {\bibinfo {volume} {82}},\ \bibinfo {pages} {124401} (\bibinfo {year} {2019})}\BibitemShut {NoStop}%
\bibitem [{\citenamefont {Kwiat}\ \emph {et~al.}(1995)\citenamefont {Kwiat}, \citenamefont {Mattle}, \citenamefont {Weinfurter}, \citenamefont {Zeilinger}, \citenamefont {Sergienko},\ and\ \citenamefont {Shih}}]{Kwiat1995}%
  \BibitemOpen
  \bibfield  {author} {\bibinfo {author} {\bibfnamefont {P.~G.}\ \bibnamefont {Kwiat}}, \bibinfo {author} {\bibfnamefont {K.}~\bibnamefont {Mattle}}, \bibinfo {author} {\bibfnamefont {H.}~\bibnamefont {Weinfurter}}, \bibinfo {author} {\bibfnamefont {A.}~\bibnamefont {Zeilinger}}, \bibinfo {author} {\bibfnamefont {A.~V.}\ \bibnamefont {Sergienko}},\ and\ \bibinfo {author} {\bibfnamefont {Y.}~\bibnamefont {Shih}},\ }\bibfield  {title} {\bibinfo {title} {{New high-intensity source of polarization-entangled photon pairs}},\ }\href@noop {} {\bibfield  {journal} {\bibinfo  {journal} {Physical Review Letters}\ }\textbf {\bibinfo {volume} {75}},\ \bibinfo {pages} {24} (\bibinfo {year} {1995})}\BibitemShut {NoStop}%
\bibitem [{\citenamefont {Mair}\ \emph {et~al.}(2001)\citenamefont {Mair}, \citenamefont {Vaziri}, \citenamefont {Weihs},\ and\ \citenamefont {Zeilinger}}]{Mair2001}%
  \BibitemOpen
  \bibfield  {author} {\bibinfo {author} {\bibfnamefont {A.}~\bibnamefont {Mair}}, \bibinfo {author} {\bibfnamefont {A.}~\bibnamefont {Vaziri}}, \bibinfo {author} {\bibfnamefont {G.}~\bibnamefont {Weihs}},\ and\ \bibinfo {author} {\bibfnamefont {A.}~\bibnamefont {Zeilinger}},\ }\bibfield  {title} {\bibinfo {title} {{Entanglement of the orbital angular momentum states of photons}},\ }\href@noop {} {\bibfield  {journal} {\bibinfo  {journal} {Nature}\ }\textbf {\bibinfo {volume} {412}} (\bibinfo {year} {2001})}\BibitemShut {NoStop}%
\bibitem [{\citenamefont {Romero}\ \emph {et~al.}(2011)\citenamefont {Romero}, \citenamefont {Leach}, \citenamefont {Jack}, \citenamefont {Dennis}, \citenamefont {Franke-Arnold}, \citenamefont {Barnett},\ and\ \citenamefont {Padgett}}]{Romero2011}%
  \BibitemOpen
  \bibfield  {author} {\bibinfo {author} {\bibfnamefont {J.}~\bibnamefont {Romero}}, \bibinfo {author} {\bibfnamefont {J.}~\bibnamefont {Leach}}, \bibinfo {author} {\bibfnamefont {B.}~\bibnamefont {Jack}}, \bibinfo {author} {\bibfnamefont {M.~R.}\ \bibnamefont {Dennis}}, \bibinfo {author} {\bibfnamefont {S.}~\bibnamefont {Franke-Arnold}}, \bibinfo {author} {\bibfnamefont {S.~M.}\ \bibnamefont {Barnett}},\ and\ \bibinfo {author} {\bibfnamefont {M.~J.}\ \bibnamefont {Padgett}},\ }\bibfield  {title} {\bibinfo {title} {{Entangled optical vortex links}},\ }\href@noop {} {\bibfield  {journal} {\bibinfo  {journal} {Physical Review Letters}\ }\textbf {\bibinfo {volume} {106}},\ \bibinfo {pages} {100407} (\bibinfo {year} {2011})}\BibitemShut {NoStop}%
\bibitem [{\citenamefont {Mclaren}\ \emph {et~al.}(2012)\citenamefont {Mclaren}, \citenamefont {Agnew}, \citenamefont {Leach}, \citenamefont {Roux}, \citenamefont {Padgett}, \citenamefont {Boyd},\ and\ \citenamefont {Forbes}}]{Mclaren2012}%
  \BibitemOpen
  \bibfield  {author} {\bibinfo {author} {\bibfnamefont {M.}~\bibnamefont {Mclaren}}, \bibinfo {author} {\bibfnamefont {M.}~\bibnamefont {Agnew}}, \bibinfo {author} {\bibfnamefont {J.}~\bibnamefont {Leach}}, \bibinfo {author} {\bibfnamefont {F.~S.}\ \bibnamefont {Roux}}, \bibinfo {author} {\bibfnamefont {M.~J.}\ \bibnamefont {Padgett}}, \bibinfo {author} {\bibfnamefont {R.~W.}\ \bibnamefont {Boyd}},\ and\ \bibinfo {author} {\bibfnamefont {A.}~\bibnamefont {Forbes}},\ }\bibfield  {title} {\bibinfo {title} {{Entangled Bessel-Gaussian beams}},\ }\href@noop {} {\bibfield  {journal} {\bibinfo  {journal} {Optics Express}\ }\textbf {\bibinfo {volume} {20}},\ \bibinfo {pages} {21} (\bibinfo {year} {2012})}\BibitemShut {NoStop}%
\bibitem [{\citenamefont {McLaren}\ \emph {et~al.}(2015)\citenamefont {McLaren}, \citenamefont {Roux},\ and\ \citenamefont {Forbes}}]{McLaren2015}%
  \BibitemOpen
  \bibfield  {author} {\bibinfo {author} {\bibfnamefont {M.~G.}\ \bibnamefont {McLaren}}, \bibinfo {author} {\bibfnamefont {F.~S.}\ \bibnamefont {Roux}},\ and\ \bibinfo {author} {\bibfnamefont {A.}~\bibnamefont {Forbes}},\ }\bibfield  {title} {\bibinfo {title} {{Realising high-dimensional quantum entanglement with orbital angular momentum}},\ }\href@noop {} {\bibfield  {journal} {\bibinfo  {journal} {South African Journal of Science}\ }\textbf {\bibinfo {volume} {111}},\ \bibinfo {pages} {1/2} (\bibinfo {year} {2015})}\BibitemShut {NoStop}%
\bibitem [{\citenamefont {Restuccia}\ \emph {et~al.}(2016)\citenamefont {Restuccia}, \citenamefont {Giovannini}, \citenamefont {Gibson},\ and\ \citenamefont {Padgett}}]{Restuccia2016}%
  \BibitemOpen
  \bibfield  {author} {\bibinfo {author} {\bibfnamefont {S.}~\bibnamefont {Restuccia}}, \bibinfo {author} {\bibfnamefont {D.}~\bibnamefont {Giovannini}}, \bibinfo {author} {\bibfnamefont {G.}~\bibnamefont {Gibson}},\ and\ \bibinfo {author} {\bibfnamefont {M.}~\bibnamefont {Padgett}},\ }\bibfield  {title} {\bibinfo {title} {{Comparing the information capacity of Laguerre–Gaussian and Hermite–Gaussian modal sets in a finite-aperture system}},\ }\href@noop {} {\bibfield  {journal} {\bibinfo  {journal} {Optics Express}\ }\textbf {\bibinfo {volume} {24}},\ \bibinfo {pages} {24} (\bibinfo {year} {2016})}\BibitemShut {NoStop}%
\bibitem [{\citenamefont {Baghdasaryan}\ and\ \citenamefont {Fritzsche}(2020)}]{Baghdasaryan2020}%
  \BibitemOpen
  \bibfield  {author} {\bibinfo {author} {\bibfnamefont {B.}~\bibnamefont {Baghdasaryan}}\ and\ \bibinfo {author} {\bibfnamefont {S.}~\bibnamefont {Fritzsche}},\ }\bibfield  {title} {\bibinfo {title} {{Enhanced entanglement from Ince-Gaussian pump beams in spontaneous parametric down-conversion}},\ }\href@noop {} {\bibfield  {journal} {\bibinfo  {journal} {Physical Review A}\ }\textbf {\bibinfo {volume} {102}},\ \bibinfo {pages} {052412} (\bibinfo {year} {2020})}\BibitemShut {NoStop}%
\bibitem [{\citenamefont {Lib}\ and\ \citenamefont {Bromberg}(2020)}]{Lib2020}%
  \BibitemOpen
  \bibfield  {author} {\bibinfo {author} {\bibfnamefont {O.}~\bibnamefont {Lib}}\ and\ \bibinfo {author} {\bibfnamefont {Y.}~\bibnamefont {Bromberg}},\ }\bibfield  {title} {\bibinfo {title} {{Spatially entangled Airy photons}},\ }\href@noop {} {\bibfield  {journal} {\bibinfo  {journal} {Optics Letters}\ }\textbf {\bibinfo {volume} {45}},\ \bibinfo {pages} {6} (\bibinfo {year} {2020})}\BibitemShut {NoStop}%
\bibitem [{\citenamefont {Valencia}\ \emph {et~al.}(2021)\citenamefont {Valencia}, \citenamefont {Srivastav}, \citenamefont {Leedumrongwatthanakun}, \citenamefont {McCutcheon},\ and\ \citenamefont {Malik}}]{Valencia2021}%
  \BibitemOpen
  \bibfield  {author} {\bibinfo {author} {\bibfnamefont {N.~H.}\ \bibnamefont {Valencia}}, \bibinfo {author} {\bibfnamefont {V.}~\bibnamefont {Srivastav}}, \bibinfo {author} {\bibfnamefont {S.}~\bibnamefont {Leedumrongwatthanakun}}, \bibinfo {author} {\bibfnamefont {W.}~\bibnamefont {McCutcheon}},\ and\ \bibinfo {author} {\bibfnamefont {M.}~\bibnamefont {Malik}},\ }\bibfield  {title} {\bibinfo {title} {{Entangled ripples and twists of light: radial and azimuthal Laguerre-Gaussian mode entanglement}},\ }\href@noop {} {\bibfield  {journal} {\bibinfo  {journal} {Journal of Optics}\ }\textbf {\bibinfo {volume} {23}},\ \bibinfo {pages} {104001} (\bibinfo {year} {2021})}\BibitemShut {NoStop}%
\bibitem [{\citenamefont {Picart}(2015)}]{Picart2015}%
  \BibitemOpen
  \bibfield  {author} {\bibinfo {author} {\bibfnamefont {P.}~\bibnamefont {Picart}},\ }\href@noop {} {\emph {\bibinfo {title} {{New techniques in digital holography}}}}\ (\bibinfo  {publisher} {Wiley},\ \bibinfo {address} {Hoboken, New Jersey},\ \bibinfo {year} {2015})\BibitemShut {NoStop}%
\bibitem [{\citenamefont {Agnew}\ \emph {et~al.}(2011)\citenamefont {Agnew}, \citenamefont {Leach}, \citenamefont {Mclaren}, \citenamefont {Roux},\ and\ \citenamefont {Boyd}}]{Agnew2011}%
  \BibitemOpen
  \bibfield  {author} {\bibinfo {author} {\bibfnamefont {M.}~\bibnamefont {Agnew}}, \bibinfo {author} {\bibfnamefont {J.}~\bibnamefont {Leach}}, \bibinfo {author} {\bibfnamefont {M.}~\bibnamefont {Mclaren}}, \bibinfo {author} {\bibfnamefont {F.~S.}\ \bibnamefont {Roux}},\ and\ \bibinfo {author} {\bibfnamefont {R.~W.}\ \bibnamefont {Boyd}},\ }\bibfield  {title} {\bibinfo {title} {{Tomography of the quantum state of photons entangled in high dimensions}},\ }\href@noop {} {\bibfield  {journal} {\bibinfo  {journal} {Physical Review A}\ }\textbf {\bibinfo {volume} {84}},\ \bibinfo {pages} {062101} (\bibinfo {year} {2011})}\BibitemShut {NoStop}%
\bibitem [{\citenamefont {Toninelli}\ \emph {et~al.}(2019)\citenamefont {Toninelli}, \citenamefont {Ndagano}, \citenamefont {Vall{\'{e}}s}, \citenamefont {Sephton}, \citenamefont {Nape}, \citenamefont {Ambrosio}, \citenamefont {Capasso}, \citenamefont {Padgett},\ and\ \citenamefont {Forbes}}]{Toninelli2019}%
  \BibitemOpen
  \bibfield  {author} {\bibinfo {author} {\bibfnamefont {E.}~\bibnamefont {Toninelli}}, \bibinfo {author} {\bibfnamefont {B.}~\bibnamefont {Ndagano}}, \bibinfo {author} {\bibfnamefont {A.}~\bibnamefont {Vall{\'{e}}s}}, \bibinfo {author} {\bibfnamefont {B.}~\bibnamefont {Sephton}}, \bibinfo {author} {\bibfnamefont {I.}~\bibnamefont {Nape}}, \bibinfo {author} {\bibfnamefont {A.}~\bibnamefont {Ambrosio}}, \bibinfo {author} {\bibfnamefont {F.}~\bibnamefont {Capasso}}, \bibinfo {author} {\bibfnamefont {M.~J.}\ \bibnamefont {Padgett}},\ and\ \bibinfo {author} {\bibfnamefont {A.}~\bibnamefont {Forbes}},\ }\bibfield  {title} {\bibinfo {title} {{Concepts in quantum state tomography and classical implementation with intense light: a tutorial}},\ }\href@noop {} {\bibfield  {journal} {\bibinfo  {journal} {Advances in Optics and Photonics}\ }\textbf {\bibinfo {volume} {11}},\ \bibinfo {pages} {1} (\bibinfo {year} {2019})}\BibitemShut {NoStop}%
\bibitem [{\citenamefont {Li}\ \emph {et~al.}(2023)\citenamefont {Li}, \citenamefont {Huang}, \citenamefont {Wang}, \citenamefont {Tu}, \citenamefont {Wang}, \citenamefont {Li},\ and\ \citenamefont {Wang}}]{Li2023}%
  \BibitemOpen
  \bibfield  {author} {\bibinfo {author} {\bibfnamefont {Y.}~\bibnamefont {Li}}, \bibinfo {author} {\bibfnamefont {S.-Y.}\ \bibnamefont {Huang}}, \bibinfo {author} {\bibfnamefont {M.}~\bibnamefont {Wang}}, \bibinfo {author} {\bibfnamefont {C.}~\bibnamefont {Tu}}, \bibinfo {author} {\bibfnamefont {X.-L.}\ \bibnamefont {Wang}}, \bibinfo {author} {\bibfnamefont {Y.}~\bibnamefont {Li}},\ and\ \bibinfo {author} {\bibfnamefont {H.-T.}\ \bibnamefont {Wang}},\ }\bibfield  {title} {\bibinfo {title} {{Two-Measurement Tomography of High-Dimensional Orbital Angular Momentum Entanglement}},\ }\href@noop {} {\bibfield  {journal} {\bibinfo  {journal} {Physical Review Letters}\ }\textbf {\bibinfo {volume} {130}},\ \bibinfo {pages} {050805} (\bibinfo {year} {2023})}\BibitemShut {NoStop}%
\bibitem [{\citenamefont {Sahoo}\ \emph {et~al.}(2020)\citenamefont {Sahoo}, \citenamefont {Chakraborti}, \citenamefont {Pati},\ and\ \citenamefont {Sinha}}]{Sahoo2020}%
  \BibitemOpen
  \bibfield  {author} {\bibinfo {author} {\bibfnamefont {S.~N.}\ \bibnamefont {Sahoo}}, \bibinfo {author} {\bibfnamefont {S.}~\bibnamefont {Chakraborti}}, \bibinfo {author} {\bibfnamefont {A.~K.}\ \bibnamefont {Pati}},\ and\ \bibinfo {author} {\bibfnamefont {U.}~\bibnamefont {Sinha}},\ }\bibfield  {title} {\bibinfo {title} {{Quantum state interferography}},\ }\href@noop {} {\bibfield  {journal} {\bibinfo  {journal} {Physical Review Letters}\ }\textbf {\bibinfo {volume} {125}},\ \bibinfo {pages} {123601} (\bibinfo {year} {2020})}\BibitemShut {NoStop}%
\bibitem [{\citenamefont {Pan}\ \emph {et~al.}(2019)\citenamefont {Pan}, \citenamefont {Xu}, \citenamefont {Kedem}, \citenamefont {Wang}, \citenamefont {Chen}, \citenamefont {Jan}, \citenamefont {Sun}, \citenamefont {Xu}, \citenamefont {Han}, \citenamefont {Li},\ and\ \citenamefont {Guo}}]{Pan2019}%
  \BibitemOpen
  \bibfield  {author} {\bibinfo {author} {\bibfnamefont {W.-W.}\ \bibnamefont {Pan}}, \bibinfo {author} {\bibfnamefont {X.-Y.}\ \bibnamefont {Xu}}, \bibinfo {author} {\bibfnamefont {Y.}~\bibnamefont {Kedem}}, \bibinfo {author} {\bibfnamefont {Q.-Q.}\ \bibnamefont {Wang}}, \bibinfo {author} {\bibfnamefont {Z.}~\bibnamefont {Chen}}, \bibinfo {author} {\bibfnamefont {M.}~\bibnamefont {Jan}}, \bibinfo {author} {\bibfnamefont {K.}~\bibnamefont {Sun}}, \bibinfo {author} {\bibfnamefont {J.-S.}\ \bibnamefont {Xu}}, \bibinfo {author} {\bibfnamefont {Y.-J.}\ \bibnamefont {Han}}, \bibinfo {author} {\bibfnamefont {C.-F.}\ \bibnamefont {Li}},\ and\ \bibinfo {author} {\bibfnamefont {G.-C.}\ \bibnamefont {Guo}},\ }\bibfield  {title} {\bibinfo {title} {{Direct Measurement of a Nonlocal Entangled Quantum State}},\ }\href@noop {} {\bibfield  {journal} {\bibinfo  {journal} {Physical Review Letters}\ }\textbf {\bibinfo {volume} {123}},\ \bibinfo {pages} {150402} (\bibinfo {year} {2019})}\BibitemShut {NoStop}%
\bibitem [{\citenamefont {Yamaguchi}\ and\ \citenamefont {Zhang}(1997)}]{Yamaguchi1997}%
  \BibitemOpen
  \bibfield  {author} {\bibinfo {author} {\bibfnamefont {I.}~\bibnamefont {Yamaguchi}}\ and\ \bibinfo {author} {\bibfnamefont {T.}~\bibnamefont {Zhang}},\ }\bibfield  {title} {\bibinfo {title} {{Phase-shifting digital holography}},\ }\href@noop {} {\bibfield  {journal} {\bibinfo  {journal} {Optics Letters}\ }\textbf {\bibinfo {volume} {22}},\ \bibinfo {pages} {16} (\bibinfo {year} {1997})}\BibitemShut {NoStop}%
\bibitem [{\citenamefont {Andersen}\ \emph {et~al.}(2019)\citenamefont {Andersen}, \citenamefont {Alperin}, \citenamefont {Voitiv}, \citenamefont {Holtzmann}, \citenamefont {Gopinath},\ and\ \citenamefont {Siemens}}]{Andersen2019}%
  \BibitemOpen
  \bibfield  {author} {\bibinfo {author} {\bibfnamefont {J.~M.}\ \bibnamefont {Andersen}}, \bibinfo {author} {\bibfnamefont {S.~N.}\ \bibnamefont {Alperin}}, \bibinfo {author} {\bibfnamefont {A.~A.}\ \bibnamefont {Voitiv}}, \bibinfo {author} {\bibfnamefont {W.~G.}\ \bibnamefont {Holtzmann}}, \bibinfo {author} {\bibfnamefont {J.~T.}\ \bibnamefont {Gopinath}},\ and\ \bibinfo {author} {\bibfnamefont {M.~E.}\ \bibnamefont {Siemens}},\ }\bibfield  {title} {\bibinfo {title} {{Characterizing vortex beams from a spatial light modulator with collinear phase-shifting holography}},\ }\href@noop {} {\bibfield  {journal} {\bibinfo  {journal} {Applied Optics}\ }\textbf {\bibinfo {volume} {58}},\ \bibinfo {pages} {2} (\bibinfo {year} {2019})}\BibitemShut {NoStop}%
\bibitem [{\citenamefont {Saleh}\ \emph {et~al.}(2000)\citenamefont {Saleh}, \citenamefont {Abouraddy}, \citenamefont {Sergienko},\ and\ \citenamefont {Teich}}]{Saleh2000}%
  \BibitemOpen
  \bibfield  {author} {\bibinfo {author} {\bibfnamefont {B.~E.}\ \bibnamefont {Saleh}}, \bibinfo {author} {\bibfnamefont {A.~F.}\ \bibnamefont {Abouraddy}}, \bibinfo {author} {\bibfnamefont {A.~V.}\ \bibnamefont {Sergienko}},\ and\ \bibinfo {author} {\bibfnamefont {M.~C.}\ \bibnamefont {Teich}},\ }\bibfield  {title} {\bibinfo {title} {{Duality between partial coherence and partial entanglement}},\ }\href@noop {} {\bibfield  {journal} {\bibinfo  {journal} {Physical Review A}\ }\textbf {\bibinfo {volume} {62}},\ \bibinfo {pages} {043816} (\bibinfo {year} {2000})}\BibitemShut {NoStop}%
\bibitem [{\citenamefont {Bouchard}\ \emph {et~al.}(2018)\citenamefont {Bouchard}, \citenamefont {Valencia}, \citenamefont {Brandt}, \citenamefont {Fickler}, \citenamefont {Huber},\ and\ \citenamefont {Malik}}]{Bouchard2018}%
  \BibitemOpen
  \bibfield  {author} {\bibinfo {author} {\bibfnamefont {F.}~\bibnamefont {Bouchard}}, \bibinfo {author} {\bibfnamefont {N.~H.}\ \bibnamefont {Valencia}}, \bibinfo {author} {\bibfnamefont {F.}~\bibnamefont {Brandt}}, \bibinfo {author} {\bibfnamefont {R.}~\bibnamefont {Fickler}}, \bibinfo {author} {\bibfnamefont {M.}~\bibnamefont {Huber}},\ and\ \bibinfo {author} {\bibfnamefont {M.}~\bibnamefont {Malik}},\ }\bibfield  {title} {\bibinfo {title} {{Measuring azimuthal and radial modes of photons}},\ }\href@noop {} {\bibfield  {journal} {\bibinfo  {journal} {Optics Express}\ }\textbf {\bibinfo {volume} {26}},\ \bibinfo {pages} {24} (\bibinfo {year} {2018})}\BibitemShut {NoStop}%
\bibitem [{\citenamefont {Taylor}(1997)}]{Taylor1997}%
  \BibitemOpen
  \bibfield  {author} {\bibinfo {author} {\bibfnamefont {J.~R.}\ \bibnamefont {Taylor}},\ }\href@noop {} {\emph {\bibinfo {title} {{An Introduction to Error Analysis: The Study of Uncertainties in Physical Measurements}}}}\ (\bibinfo  {publisher} {University Science Books},\ \bibinfo {address} {Melville, New York},\ \bibinfo {year} {1997})\BibitemShut {NoStop}%
\bibitem [{\citenamefont {Padgett}\ and\ \citenamefont {Courtial}(1999)}]{Padgett1999}%
  \BibitemOpen
  \bibfield  {author} {\bibinfo {author} {\bibfnamefont {M.~J.}\ \bibnamefont {Padgett}}\ and\ \bibinfo {author} {\bibfnamefont {J.}~\bibnamefont {Courtial}},\ }\bibfield  {title} {\bibinfo {title} {{Poincar{\'{e}}-sphere equivalent for light beams containing orbital angular momentum}},\ }\href@noop {} {\bibfield  {journal} {\bibinfo  {journal} {Optics Letters}\ }\textbf {\bibinfo {volume} {24}},\ \bibinfo {pages} {7} (\bibinfo {year} {1999})}\BibitemShut {NoStop}%
\bibitem [{\citenamefont {Galvez}\ \emph {et~al.}(2003)\citenamefont {Galvez}, \citenamefont {Crawford}, \citenamefont {Sztul}, \citenamefont {Pysher}, \citenamefont {Haglin},\ and\ \citenamefont {Williams}}]{Galvez2003}%
  \BibitemOpen
  \bibfield  {author} {\bibinfo {author} {\bibfnamefont {E.~J.}\ \bibnamefont {Galvez}}, \bibinfo {author} {\bibfnamefont {P.~R.}\ \bibnamefont {Crawford}}, \bibinfo {author} {\bibfnamefont {H.~I.}\ \bibnamefont {Sztul}}, \bibinfo {author} {\bibfnamefont {M.~J.}\ \bibnamefont {Pysher}}, \bibinfo {author} {\bibfnamefont {P.~J.}\ \bibnamefont {Haglin}},\ and\ \bibinfo {author} {\bibfnamefont {R.~E.}\ \bibnamefont {Williams}},\ }\bibfield  {title} {\bibinfo {title} {{Geometric Phase Associated with Mode Transformations of Optical Beams Bearing Orbital Angular Momentum}},\ }\href@noop {} {\bibfield  {journal} {\bibinfo  {journal} {Physical Review Letters}\ }\textbf {\bibinfo {volume} {90}},\ \bibinfo {pages} {20} (\bibinfo {year} {2003})}\BibitemShut {NoStop}%
\bibitem [{\citenamefont {Lusk}\ \emph {et~al.}(2022)\citenamefont {Lusk}, \citenamefont {Voitiv}, \citenamefont {Zhu},\ and\ \citenamefont {Siemens}}]{Lusk2022}%
  \BibitemOpen
  \bibfield  {author} {\bibinfo {author} {\bibfnamefont {M.~T.}\ \bibnamefont {Lusk}}, \bibinfo {author} {\bibfnamefont {A.~A.}\ \bibnamefont {Voitiv}}, \bibinfo {author} {\bibfnamefont {C.}~\bibnamefont {Zhu}},\ and\ \bibinfo {author} {\bibfnamefont {M.~E.}\ \bibnamefont {Siemens}},\ }\bibfield  {title} {\bibinfo {title} {{The Anatomy of Geometric Phase for an Optical Vortex Transiting a Lens}},\ }\href@noop {} {\bibfield  {journal} {\bibinfo  {journal} {Physical Review A}\ }\textbf {\bibinfo {volume} {105}},\ \bibinfo {pages} {052211} (\bibinfo {year} {2022})}\BibitemShut {NoStop}%
\bibitem [{\citenamefont {Voitiv}\ \emph {et~al.}(2022)\citenamefont {Voitiv}, \citenamefont {Lusk},\ and\ \citenamefont {Siemens}}]{Voitiv2022}%
  \BibitemOpen
  \bibfield  {author} {\bibinfo {author} {\bibfnamefont {A.~A.}\ \bibnamefont {Voitiv}}, \bibinfo {author} {\bibfnamefont {M.~T.}\ \bibnamefont {Lusk}},\ and\ \bibinfo {author} {\bibfnamefont {M.~E.}\ \bibnamefont {Siemens}},\ }\bibfield  {title} {\bibinfo {title} {{Tilted Poincar{\'{e}} sphere geodesics}},\ }\href@noop {} {\bibfield  {journal} {\bibinfo  {journal} {Optics Letters}\ }\textbf {\bibinfo {volume} {47}},\ \bibinfo {pages} {5} (\bibinfo {year} {2022})}\BibitemShut {NoStop}%
\bibitem [{\citenamefont {Franke-Arnold}\ \emph {et~al.}(2002)\citenamefont {Franke-Arnold}, \citenamefont {Barnett}, \citenamefont {Padgett},\ and\ \citenamefont {Allen}}]{Franke-Arnold2002}%
  \BibitemOpen
  \bibfield  {author} {\bibinfo {author} {\bibfnamefont {S.}~\bibnamefont {Franke-Arnold}}, \bibinfo {author} {\bibfnamefont {S.~M.}\ \bibnamefont {Barnett}}, \bibinfo {author} {\bibfnamefont {M.~J.}\ \bibnamefont {Padgett}},\ and\ \bibinfo {author} {\bibfnamefont {L.}~\bibnamefont {Allen}},\ }\bibfield  {title} {\bibinfo {title} {{Two-photon entanglement of orbital angular momentum states}},\ }\href@noop {} {\bibfield  {journal} {\bibinfo  {journal} {Physical Review A}\ }\textbf {\bibinfo {volume} {65}},\ \bibinfo {pages} {033823} (\bibinfo {year} {2002})}\BibitemShut {NoStop}%
\bibitem [{\citenamefont {Bornman}\ \emph {et~al.}(2021)\citenamefont {Bornman}, \citenamefont {Buono}, \citenamefont {Lovemore},\ and\ \citenamefont {Forbes}}]{Bornman2021}%
  \BibitemOpen
  \bibfield  {author} {\bibinfo {author} {\bibfnamefont {N.}~\bibnamefont {Bornman}}, \bibinfo {author} {\bibfnamefont {W.~T.}\ \bibnamefont {Buono}}, \bibinfo {author} {\bibfnamefont {M.}~\bibnamefont {Lovemore}},\ and\ \bibinfo {author} {\bibfnamefont {A.}~\bibnamefont {Forbes}},\ }\bibfield  {title} {\bibinfo {title} {{Optimal Pump Shaping for Entanglement Control in Any Countable Basis}},\ }\href@noop {} {\bibfield  {journal} {\bibinfo  {journal} {Advanced Quantum Technologies}\ }\textbf {\bibinfo {volume} {4}},\ \bibinfo {pages} {2100066} (\bibinfo {year} {2021})}\BibitemShut {NoStop}%
\bibitem [{\citenamefont {Hiekkam\"aki}\ \emph {et~al.}(2022)\citenamefont {Hiekkam\"aki}, \citenamefont {Barros}, \citenamefont {Ornigotti},\ and\ \citenamefont {Fickler}}]{Hiekkamaki2021}%
  \BibitemOpen
  \bibfield  {author} {\bibinfo {author} {\bibfnamefont {M.}~\bibnamefont {Hiekkam\"aki}}, \bibinfo {author} {\bibfnamefont {R.~F.}\ \bibnamefont {Barros}}, \bibinfo {author} {\bibfnamefont {M.}~\bibnamefont {Ornigotti}},\ and\ \bibinfo {author} {\bibfnamefont {R.}~\bibnamefont {Fickler}},\ }\bibfield  {title} {\bibinfo {title} {{Observation of the quantum Gouy phase}},\ }\href@noop {} {\bibfield  {journal} {\bibinfo  {journal} {Nature Photonics}\ }\textbf {\bibinfo {volume} {16}},\ \bibinfo {pages} {828} (\bibinfo {year} {2022})}\BibitemShut {NoStop}%
\bibitem [{\citenamefont {Hong}\ \emph {et~al.}(2021)\citenamefont {Hong}, \citenamefont {ur~Rehman}, \citenamefont {Kim}, \citenamefont {Cho}, \citenamefont {Lee}, \citenamefont {Jung}, \citenamefont {Moon}, \citenamefont {Han},\ and\ \citenamefont {Lim}}]{Hong2021}%
  \BibitemOpen
  \bibfield  {author} {\bibinfo {author} {\bibfnamefont {S.}~\bibnamefont {Hong}}, \bibinfo {author} {\bibfnamefont {J.}~\bibnamefont {ur~Rehman}}, \bibinfo {author} {\bibfnamefont {Y.-S.}\ \bibnamefont {Kim}}, \bibinfo {author} {\bibfnamefont {Y.-W.}\ \bibnamefont {Cho}}, \bibinfo {author} {\bibfnamefont {S.-W.}\ \bibnamefont {Lee}}, \bibinfo {author} {\bibfnamefont {H.}~\bibnamefont {Jung}}, \bibinfo {author} {\bibfnamefont {S.}~\bibnamefont {Moon}}, \bibinfo {author} {\bibfnamefont {S.-W.}\ \bibnamefont {Han}},\ and\ \bibinfo {author} {\bibfnamefont {H.-T.}\ \bibnamefont {Lim}},\ }\bibfield  {title} {\bibinfo {title} {{Quantum enhanced multiple-phase estimation with multi-mode N00N states}},\ }\href@noop {} {\bibfield  {journal} {\bibinfo  {journal} {Nature Communications}\ }\textbf {\bibinfo {volume} {12}},\ \bibinfo {pages} {5211} (\bibinfo {year} {2021})}\BibitemShut {NoStop}%
\bibitem [{\citenamefont {Nielsen}\ \emph {et~al.}(2023)\citenamefont {Nielsen}, \citenamefont {Neergaard-Nielsen}, \citenamefont {Gehring},\ and\ \citenamefont {Andersen}}]{Nielsen2023}%
  \BibitemOpen
  \bibfield  {author} {\bibinfo {author} {\bibfnamefont {J.~A.}\ \bibnamefont {Nielsen}}, \bibinfo {author} {\bibfnamefont {J.~S.}\ \bibnamefont {Neergaard-Nielsen}}, \bibinfo {author} {\bibfnamefont {T.}~\bibnamefont {Gehring}},\ and\ \bibinfo {author} {\bibfnamefont {U.~L.}\ \bibnamefont {Andersen}},\ }\bibfield  {title} {\bibinfo {title} {{Deterministic Quantum Phase Estimation beyond N00N States}},\ }\href@noop {} {\bibfield  {journal} {\bibinfo  {journal} {Physical Review Letters}\ }\textbf {\bibinfo {volume} {130}},\ \bibinfo {pages} {123603} (\bibinfo {year} {2023})}\BibitemShut {NoStop}%
\end{thebibliography}%

\pagebreak
\widetext
\begin{center}
\textbf{\large Supplementary Document: \\ Phase-resolved measurement of entangled states via common-path interferometry}

\bigskip

Here we provide experimental methods for the results presented in the manuscript, along with additional applications of the new technique.
\end{center}

\setcounter{equation}{0}
\setcounter{figure}{0}
\setcounter{table}{0}
\setcounter{page}{1}
\makeatletter
\renewcommand{\theequation}{S.\arabic{equation}}
\renewcommand{\thefigure}{S.\arabic{figure}}

\section{S.1. Experimental Schematic and Details}

\begin{center}
    \includegraphics[width=.99\linewidth]{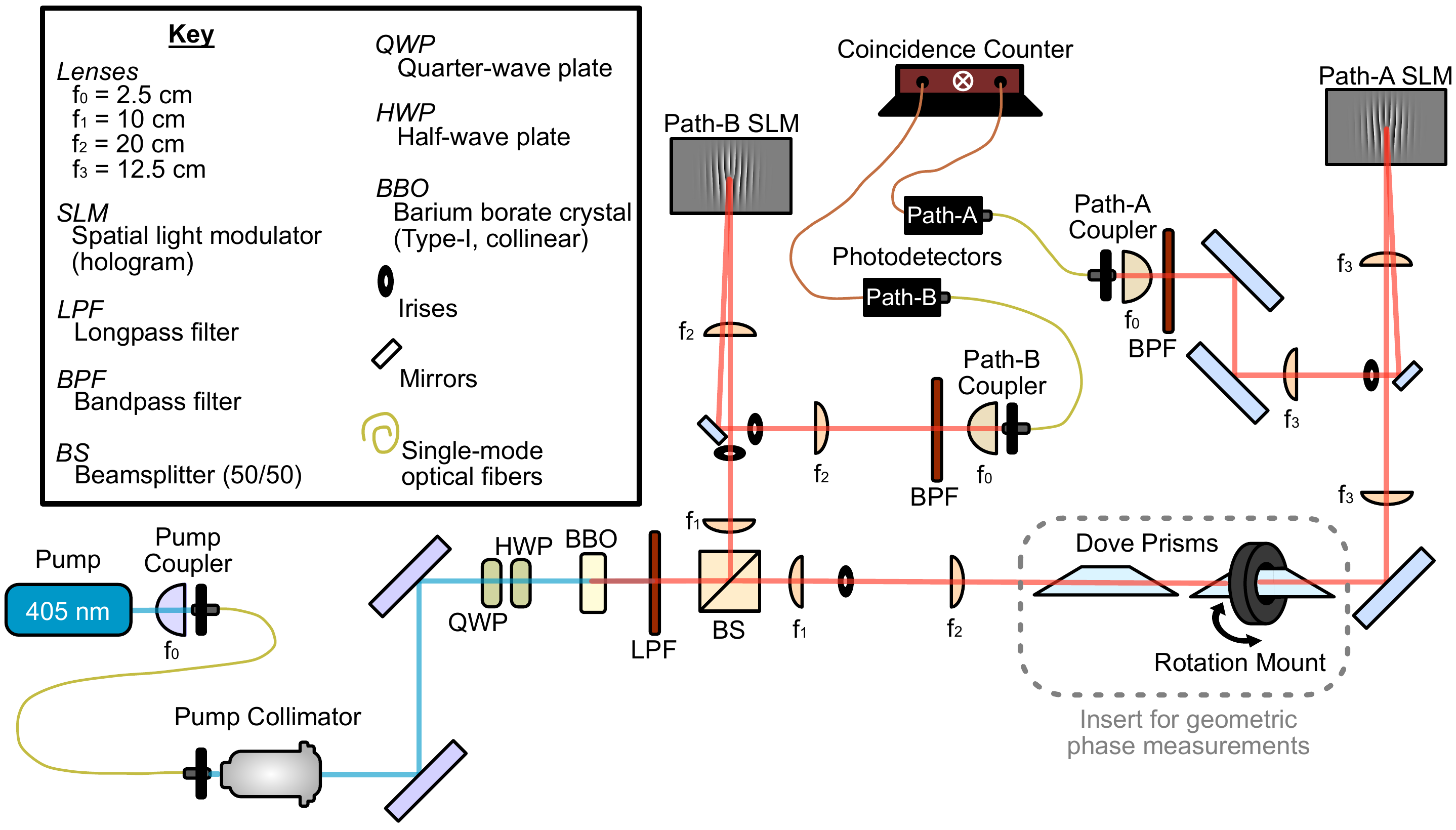}
\captionof{figure}{Full Schematic. See body of text for explanation of components.}
    \label{fig:fullschematic}
\end{center}

The experimental data presented in the manuscript were acquired using the setup depicted in Fig. \ref{fig:fullschematic}. The pump beam was a continuous wavelength diode laser (unbranded, model FA02), which we measured to have wavelength $\lambda = 406$ nm. It was coupled into a single mode optical fiber and the output was collimated to obtain a Gaussian mode to pump the nonlinear crystal. We used a $5$ mm-thick beta barium-borate (BBO) crystal cut for Type-I SPDC with zero degree opening angle (Newlight Photonics part number NCBBO5500-405(I)-HA0). Before incidence on the BBO, a quarter-wave plate and half-wave plate were used to tune the polarization of the pump photon such that the polarizations of the down-converted photons (which are identical to each other and both perpendicular to that of the pump) matched the most efficient polarization to use with the spatial light modulators (SLMs). In turn, the orientation of the BBO was aligned for maximum efficiency of down-converted photons in the zero-degree geometry. The pump beam was collimated with a waist of $0.87$ mm. The power of the pump incident on the BBO was approximately $65$ mW. After the BBO, a longpass filter (Thorlabs, cut-on wavelength of $750$ nm) was used to block the pump beam. From this point, all optics were ThorLabs B-coated or NIR-enhanced dielectric mirrors to improve efficiency for the near-infrared $812$ nm down-converted photons. A $50/50$ beamsplitter separated the photons into path-$A$ (transmitted) and path-$B$ (reflected) branches. The measured ratio was closer to $55/45$.

For Fig. 2 of the manuscript (no Dove prisms included), the output of the BBO was $4f$-imaged towards the respective measurement SLMs (each a Holoeye Pluto-2.1-NIR-118). These SLMs are reflection-based and displayed blazed diffraction gratings. The gratings contained digital holograms of various LG modes (including superpositions with the phase-stepped reference Gaussian modes), as described in the manuscript. From the SLMs, the first-order diffracted mode in each arm was $4f$-imaged into single-mode optical fiber (SMF) coupling-devices comprised of lenses in front of optical fiber mounts. The result is a projection of the mode onto the measurement operator specified by the conjugate of the mode programmed on the SLM.

Bandpass filters (Thorlabs, $800\pm20$ nm) were placed in front of each device to further limit the light to capture the $812$ nm photons of interest. The optical fibers for signal and idler each fed into photon counting modules, Excelitas SPCM-AQRH-14-FC. The electronic outputs of the photon counters were then connected to an ID Quantique ID900 Time Controller, which measured the coincidence rates presented throughout the manuscript. We used a coincidence measurement window of $3$ ns, and each data point consisted of five coincidence measurements, each collected for ten seconds.

To measure the geometric phase, as presented in Fig. 4 of the manuscript, two Dove prisms were inserted into the $A$-path of the setup at the $4f$-image plane of the BBO output, as depicted in the gray dashed box. The second Dove prism was rotated (to vary the misorientation angle $\eta$) using a motorized rotation stage, Thorlabs PRM1Z8 driven by Thorlabs KDC101.

The blazed gratings on each measurement SLM consisted of amplitude-modulated holograms of Laguerre-Gaussian (LG) modes at longitudinal distance $z=0$ and with radial charge $p=0$, expressed as
\begin{equation} \label{LG}
    E_{\mathrm{LG}}(m,x,y) = \sqrt{\frac{2}{\pi |m|!}} e^{-(x^2-y^2)/w_0^2} \left(\frac{\sqrt{2}}{w_0} \sqrt{x^2 + y^2}\right)^{|m|} e^{i \, m \, \arctan{(y/x)}},
\end{equation}
for Gaussian waist $w_0 = 2 \, \mathrm{mm}$. To measure the phase with the methods of a Gaussian ancillary reference mode, the holograms were updated to show Eqn. \ref{LG} superposed with a phase-shifted Gaussian mode (the shift was only applied to the signal SLM, as mentioned in the manuscript),
\begin{equation} \label{G}
    E_{\mathrm{G}}(\varphi_{\mathrm{ref}},x,y) = \sqrt{\frac{2}{\pi}} e^{-(x^2-y^2)/w_0^2} e^{i \, \varphi_{\mathrm{ref}}},
\end{equation}
such that we measured on the path-$A$ and path-$B$ SLMs, respectively:
\begin{gather}
    \mathrm{SLM}_{A} = d_m E_{\mathrm{LG}}(\pm m,x,y) + d_0 E_{\mathrm{G}}(0,x,y),\\
    \mathrm{SLM}_{B} = d_m E_{\mathrm{LG}}(\mp m,x,y) + d_0 E_{\mathrm{G}}(\varphi_{\mathrm{ref}},x,y).
\end{gather}
The coefficients, $d_m$ and $d_0$, were empirically determined to give the best interference contrast for the measurements. For first-order modes, we used $d_1 = 60\%$ and $d_0 = 40\%$; for second-order modes, $d_2 = 85\%$ and $d_0 = 15\%$. These values were consistent with the ratio of amplitudes between the modes measured in the intensity coincidence spectrum also reported here. To phase-resolve the coincidence spectra, four separate measurements were then made by phase-stepping the signal SLM with $\varphi_{\mathrm{ref}} = 0, \pi/2, \pi,$ and $3\pi/2$, to be used for phase-shifting digital holograph (PSDH).

Lastly, the amplitude-modulated blazed gratings (holograms, H) were constructed by interfering the above fields (SLM) with a plane wave, according to:
\begin{equation}
    \mathrm{H}(x,y) = \left[ \mathrm{arg} \left( \mathrm{SLM} \times e^{i \, k_g \, x} \right) \times \mathrm{abs} \left( \mathrm{SLM} \right) \right] \, \mathrm{mod} \, 2\pi,
\end{equation}
where the wavenumber, $k_g = 2\pi /N$, defines the fringe separation of the grating; we used $N = 16$, determined experimentally by balancing diffraction efficiency with a convenient diffraction angle.

For the measurements presented in the manuscript, the LG holograms had a waist of $w_0 = 2$ mm; this gave the highest level of coincidence readings for our setup. Typical raw coincidence readings (before normalizing as described in the manuscript) were approximately $1450$ coincidences per 10 seconds, for measuring basic $m=0$ correlations (the brightest signal).

\section{S.2. Schematic Representation of PSDH with Coincidences}

In Fig. \ref{fig:schematicpsdh} we provide an illustrative example of performing the phase calculation by making four phase-stepped coincidences and combining them in the final equation for phase-shifting digital holography (PSDH) \cite{Yamaguchi1997}. The gratings have been scaled up in physical dimension to make for easier reading here; the coincidence measurements written are data from experimental results presented in Fig. 2 of the manuscript. All coincidence measurements are the average of five measurements each acquired for 10 seconds, and errors are the first standard deviations of these measurements. The error for phase was calculated by propagating the errors for the four different coincidence measurements required to compute it \cite{Taylor1997}.

\begin{center}
    \includegraphics[width=.9\linewidth]{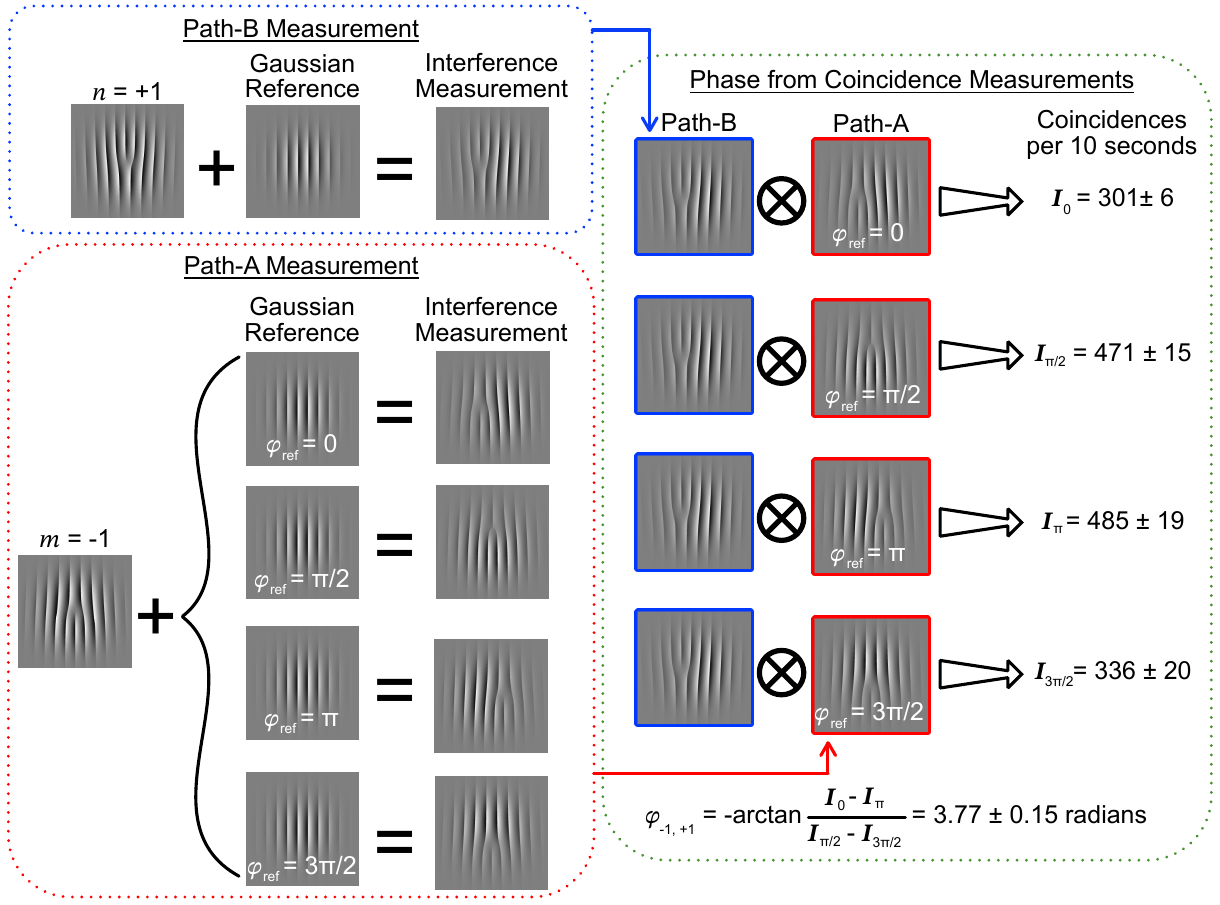}
    \captionof{figure}{Schematic depiction of measuring phase from four phase-stepped coincidence measurements.}
    \label{fig:schematicpsdh}
\end{center}

\section{S.3. Tabulated Data for Fig. 2 (a)}

To accompany the visual representation of the data shown in Fig. 2 (a) of the manuscript, Table \ref{tab:oam2} summarizes the numerical values of the diagonal elements. All coincidence measurements are the average of five measurements each acquired for 10 seconds, and errors are the first standard deviations of these measurements. Coincidences are normalized such that the sum of all elements in Fig. 2 (a) is unity. The corresponding phase measurements are in radians, modulo $2\pi$, with the phase of $\ket{0,0}$ set to zero, by definition as that is the reference mode. Errors for phase were calculated by propagating the errors for the four different coincidence measurements required to compute the phase \cite{Taylor1997}.

\begin{center}
    \begin{tabular}{c  |  c  |  c } 
    $\ket{m}_A \ket{n}_B$ & Coincidences ($a^2$) & Phases ($\varphi$) \\ [0.5ex] 
    \hline\hline
    $\ket{+2,-2}$ & $0.051 \pm 0.004$ & $1.16 \pm 0.10$ \\
    \hline
    $\ket{+1,-1}$ & $0.205 \pm 0.004$ & $3.77 \pm 0.15$ \\
    \hline
    $\ket{0,0}$ & $0.298 \pm 0.006$ & $0$ \\
    \hline
    $\ket{-1,+1}$ & $0.197 \pm 0.007$ & $3.78 \pm 0.12$ \\
    \hline
    $\ket{-2,+2}$ & $0.047 \pm 0.004$ & $1.13 \pm 0.10$ \\ [0.5ex] 
    \end{tabular}
    \captionof{table}{Diagonal-element data of Fig. 2 (a) of the manuscript.}
    \label{tab:oam2}
\end{center}

\section{S.4. Density Matrix Representation of Fig. 2 (a)}

As a demonstration of reconstructing the quantum state using all experimentally-measured values, we plot here the density matrix representations (real and imaginary) of the measured state of Fig. 2 (a), with the ancillary reference mode, $\ket{0,0}$, excluded. This is shown in Fig. \ref{fig:densitymatrix}. Plotting the full quantum state in this way shows that the quantum system is real-valued, despite the non-zero phases measured. This is because the Gouy phases measured are constant between two modes of equal $|m|$, whereas only relative phase shifts will manifest as imaginary-part contributions to the state operator in anti-diagonal elements. This is shown below in \S S.6 for the geometric-phase-shifted states.

\begin{center}
    \includegraphics[width=.5\linewidth]{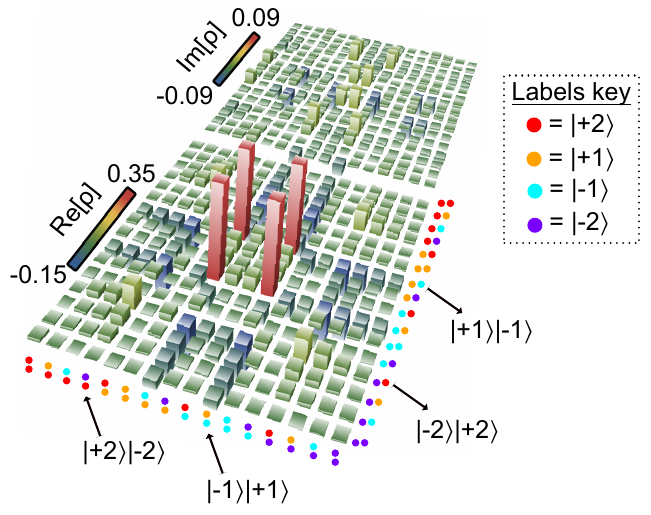}
    \captionof{figure}{Density matrix representations (real and imaginary) of the data of Fig. 2 (a) in the manuscript. Axes of different $\ket{m}_A \ket{n}_B$ states are labelled with colored dots, according to the key.}
    \label{fig:densitymatrix}
\end{center}

The density matrix (state operator) for pure states is defined as,
\begin{equation} \label{rho}
    \rho = \ket{\Psi} \bra{\Psi}.
\end{equation}
To calculate this, we first construct the experimental state as in  Eqn. 9, repeated here for convenience,
\begin{equation} \label{reconstructed}
    \ket{\Psi_{\mathrm{exp}}} = \sum_{m,n \not= 0} a_{m,n} e^{i \varphi_{m,n}} \ket{m}_A \ket{n}_B,
\end{equation}
using the amplitude and phase measurements of each $m, n$ shown in Fig. 2. These complex coefficients are direct experimental measurements which are assigned to the following vector definitions:

\begin{gather*}
    \ket{+2,+2} = (1,0,0,0,0,0,0,0,0,0,0,0,0,0,0,0),\\
    \ket{+2,+1} = (0,1,0,0,0,0,0,0,0,0,0,0,0,0,0,0),\\
    \ket{+2,-1} = (0,0,1,0,0,0,0,0,0,0,0,0,0,0,0,0),\\
    \ket{+2,-2} = (0,0,0,1,0,0,0,0,0,0,0,0,0,0,0,0),\\
    \ket{+1,+2} = (0,0,0,0,1,0,0,0,0,0,0,0,0,0,0,0),\\
    \ket{+1,+1} = (0,0,0,0,0,1,0,0,0,0,0,0,0,0,0,0),\\
    \ket{+1,-1} = (0,0,0,0,0,0,1,0,0,0,0,0,0,0,0,0),\\
    \ket{+1,-2} = (0,0,0,0,0,0,0,1,0,0,0,0,0,0,0,0),\\
    \ket{-1,+2} = (0,0,0,0,0,0,0,0,1,0,0,0,0,0,0,0),\\
    \ket{-1,+1} = (0,0,0,0,0,0,0,0,0,1,0,0,0,0,0,0),\\
    \ket{-1,-1} = (0,0,0,0,0,0,0,0,0,0,1,0,0,0,0,0),\\
    \ket{-1,-2} = (0,0,0,0,0,0,0,0,0,0,0,1,0,0,0,0),\\
    \ket{-2,+2} = (0,0,0,0,0,0,0,0,0,0,0,0,1,0,0,0),\\
    \ket{-2,+1} = (0,0,0,0,0,0,0,0,0,0,0,0,0,1,0,0),\\
    \ket{-2,-1} = (0,0,0,0,0,0,0,0,0,0,0,0,0,0,1,0),\\
    \ket{-2,-2} = (0,0,0,0,0,0,0,0,0,0,0,0,0,0,0,1).
\end{gather*}

With the full 16-term state constructed, the matrix multiplication of \ref{rho} is performed and the results are plotted in Fig. \ref{fig:densitymatrix}.

\section{S.5. Larger-spectrum Phase Measurements}

As a further test of our phase-resolving measurements, we adjusted the measurement LG modes on the SLMs to have a waist of $w_0 = 1$ mm. This was a less-optimized condition, but allowed for OAM measurements up to $m=3$ (we increased the spiral bandwidth \cite{Yao2011}). The phase-resolved coincidence spectrum is shown below in Fig. \ref{fig:oam3} (a), and (b) plots the phase values of the main diagonal. In (b), $2\pi$ is added to the $m=3$ phase measurements to show that the modes carry a phase proportional to a phase difference of $C$, attributed to the Gouy phase, as discussed in the manuscript. For these parameters, $C = -2.91$ radians, as observed from the diagonal elements' data listed in Table \ref{tab:oam3} and finding the slope connecting the points in Fig. \ref{fig:oam3} (b), as discussed in the manuscript for Fig. 2 (b). The data follows the same methods and format as described in the sections above. This value for $C$ differs from the value calculated from the spectrum in Fig. 2 of the manuscript. There, we argued that one would expect the Gouy phase to be zero for an ideal optical system with no misalignment of the focal planes of the lenses used in the schematic. These data in Fig. \ref{fig:oam3} were measured using the same experimental setup as in Fig. 2 of the manuscript (thus the same misalignment was present) with the only difference being the size of the waist measured on the SLMs. Because the Gouy phase scales with beam waist, we thus also measured different deviations in the measured phase shifts of the two different sets of data.

\begin{center}
    \begin{tabular}{c  |  c  |  c } 
    $\ket{m}_A \ket{n}_B$ & Coincidences ($a^2$) & Phases ($\varphi$) \\ [0.5ex] 
    \hline\hline
    $\ket{+3,-3}$ & $0.074 \pm 0.002$ & $-2.37 \pm 0.10$ \\ 
    \hline
    $\ket{+2,-2}$ & $0.117 \pm 0.001$ & $0.49 \pm 0.02$ \\
    \hline
    $\ket{+1,-1}$ & $0.169 \pm 0.005$ & $3.58 \pm 0.18$ \\
    \hline
    $\ket{-1,+1}$ & $0.171 \pm 0.006$ & $3.63 \pm 0.18$ \\
    \hline
    $\ket{-2,+2}$ & $0.124 \pm 0.004$ & $0.68 \pm 0.03$ \\
    \hline
    $\ket{-3,+3}$ & $0.076 \pm 0.002$ & $-2.43 \pm 0.11$ \\ [0.5ex] 
    \end{tabular}
    \captionof{table}{Diagonal-element data of Fig. \ref{fig:oam3}.}
    \label{tab:oam3}
\end{center}

\begin{center}
    \includegraphics[width=.8\linewidth]{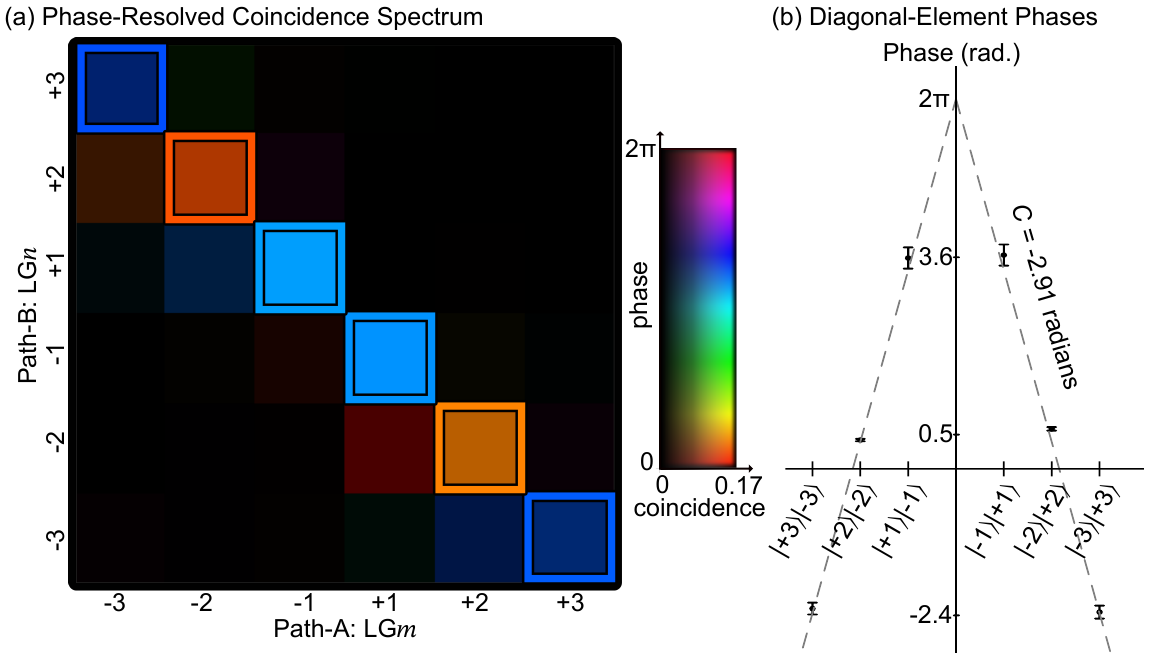}
    \captionof{figure}{(a) Phase-resolved coincidence spectrum up to $m=3$. As in Fig. 2 (a) of the manuscript, all phases are reported modulus $2\pi$. Main-diagonal elements are boxed in their corresponding color of the phase legend. (b) Phases of the four main-diagonal elements. To illustrate that the phases are scaling by a factor of $C = -2.91$ radians, the $m=3$ data from (a) is decremented by $2\pi$ (hence the y-axis extends below $0$).}
    \label{fig:oam3}
\end{center}

\section{S.6. Density Matrix Representation of Geometric-Phase-Shifted States}

To accompany the phase shift measurements of Fig. 3 (b) in the manuscript, we also provide here density matrix plots of a few reconstructed quantum states (following Eqn. \ref{reconstructed} and the methods of \S S.4), shown in Fig. \ref{fig:geophasedenmat}. This application is another way to visualize the phase-shifted quantum states by observing the varied increase/decrease of the off-diagonal elements of the density matrices for different phase accumulations---in particular, observe the swapping of signs of the off-diagonal elements in the imaginary part for the first two examples showing the geometric phase increase from $\pi/3$ to $2\pi/3$.

\begin{center}
    \includegraphics[width=.5\linewidth]{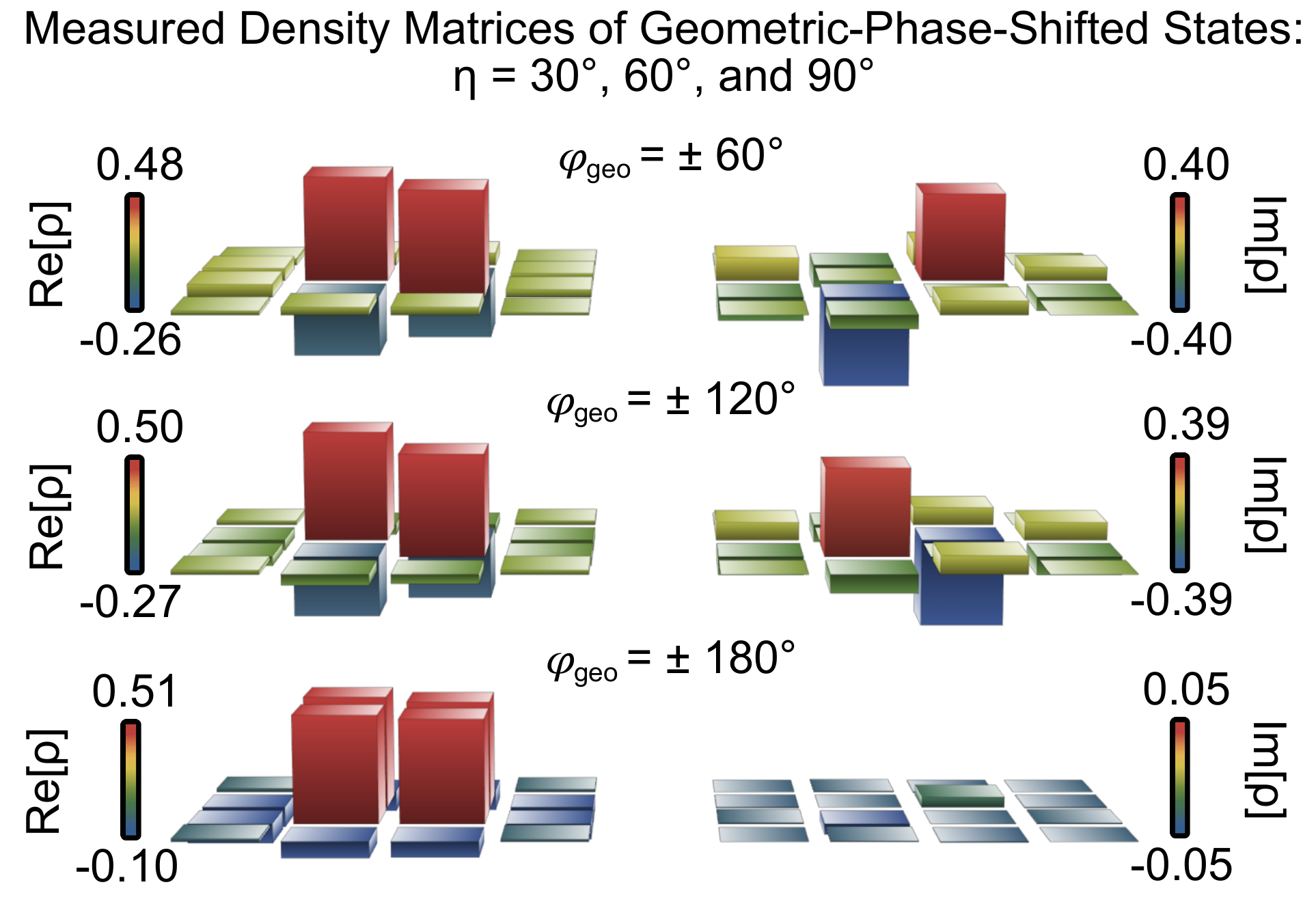}
    \captionof{figure}{Density matrices for three of the phase-shifted quantum states corresponding to data presented in Fig. 3 (b) of the manuscript.}
    \label{fig:geophasedenmat}
\end{center}

For definiteness, the vector definitions used to calculate the reconstructed matrices are:

\begin{gather*}
    \ket{+1,+1} = (1,0,0,0), \\
    \ket{+1,-1} = (0,1,0,0), \\
    \ket{-1,+1} = (0,0,1,0), \\
    \ket{-1,-1} = (0,0,0,1).
\end{gather*}

\section{S.7. Ancillary-mode Reference-stepping Path-B instead of Path-A}

As a final insight, we confirmed with the geometric phase experiment that our phase resolution method works for phase shifting either the $A$-path or $B$-path SLMs. To test this, we performed separate phase measurements for each geometric phase setting: one with phase measurement on the path-$A$ SLM and another with phase measurement on the path-$B$ SLM. Results are presented in Fig. \ref{fig:stepboth}, where the dark blue and orange data are identical to that shown in Fig. 4 (b) of the manuscript (namely, the signal SLM was phase-stepped), and the cyan and red data were measured by instead phase-stepping the path-$B$ SLM. The data shows that although the Dove prisms only act on the $A$-path, the nature of entanglement allows the geometric phase information to be extracted by phase-shifting the $B$-path measurement device. This is consistent with the phase measurements being on the product state for each component. Throughout the manuscript, arbitrary preference was assigned to performing the phase-shifting on the path-$B$ SLM.

\begin{center}
    \includegraphics[width=.5\linewidth]{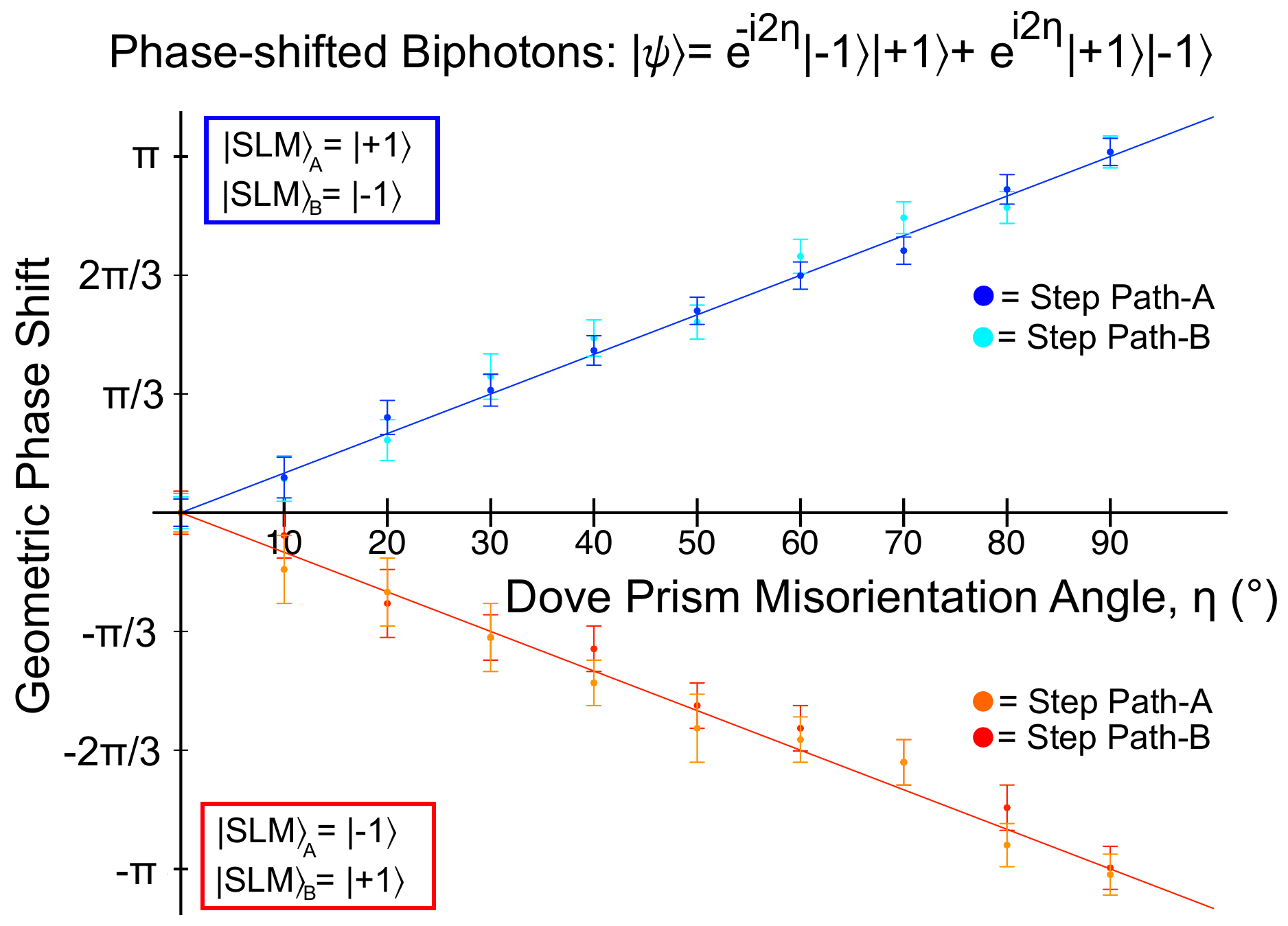}
    \captionof{figure}{Tests of phase-stepping the path-$B$ SLM, rather than the path-$A$ SLM, for the geometric phase experiment.}
    \label{fig:stepboth}
\end{center}

\newpage

\section{S.8. Full Schematic for Pump-Structured Experiment}

Full schematic used to perform the pump-structured experiment to generate the data seen in Fig. 4 of the manuscript. For full details on the different components, refer to \S S.1. The salient differences with this schematic, compared to the first, are the presence of the pump-SLM and the complete absence of the Dove prisms.

\begin{center}
    \includegraphics[width=.99\linewidth]{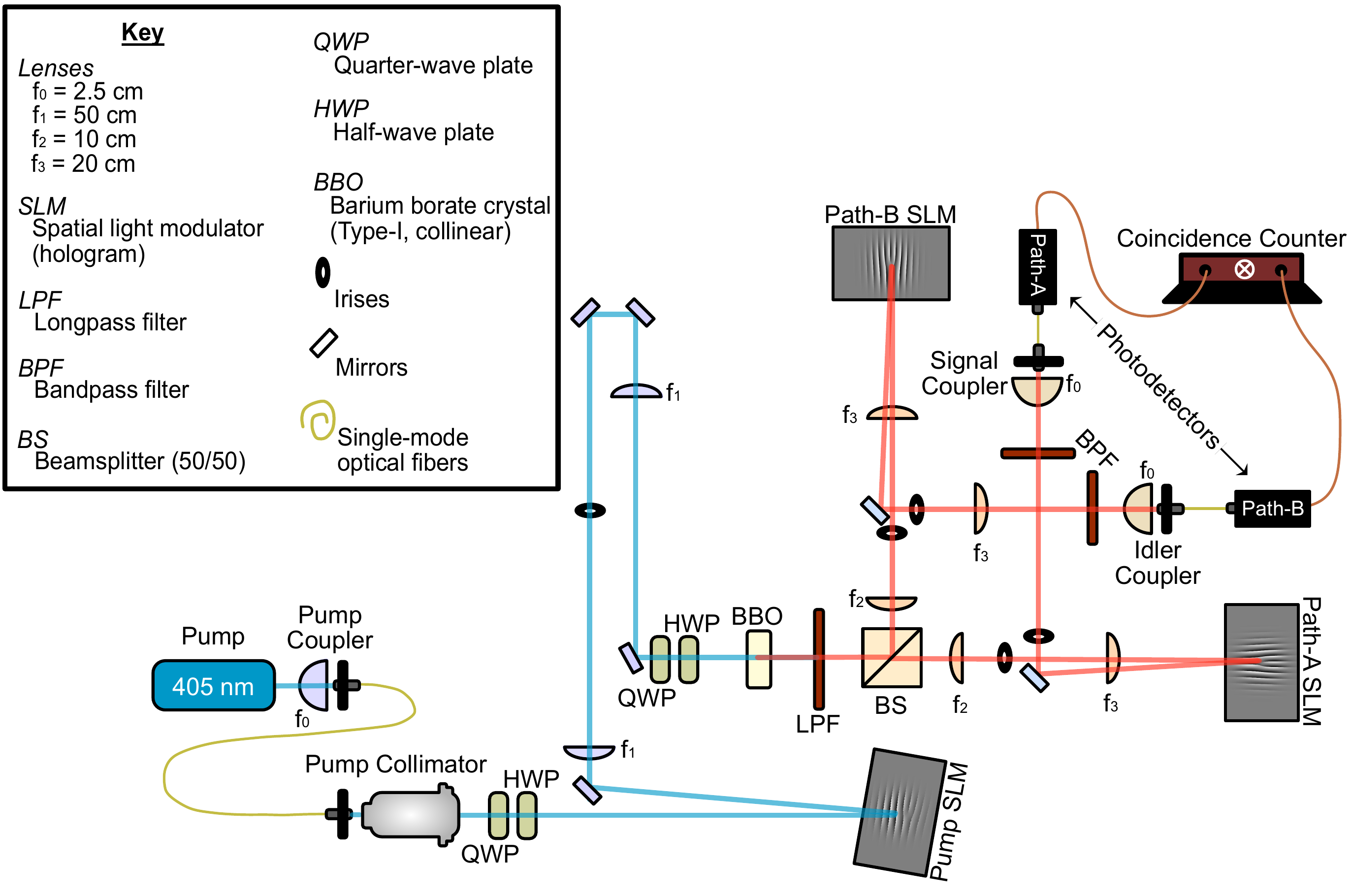}
\captionof{figure}{Full schematic including structuring of the pump beam. See text for explanation of components.}
    \label{fig:fullschematic}
\end{center}

\end{document}